\documentclass[twocolumn,showpacs,aps,prd,floatfix]{revtex4}

\usepackage{graphicx}
\usepackage{dcolumn}
\usepackage{amsmath}
\usepackage{epsfig}
\usepackage{dcolumn}
\usepackage{natbib}
\usepackage{psfrag}

\newcolumntype{.}{D{.}{.}{-1}}
\newcolumntype{,}[1]{D{.}{.}{#1}}
\newcolumntype{p}{D{\%}{\%}{3}}
\newcolumntype{a}{D{.}{\to}{-1}}

\RequirePackage{xspace}

\usepackage{relsize}
\def\babar{\mbox{\slshape B\kern-0.1em{\smaller A}\kern-0.1em
    B\kern-0.1em{\smaller A\kern-0.2em R}}}

\def\epem       {\ensuremath{e^+e^-}\xspace}

\def\W      {\ensuremath{W}\xspace}

\def\qqbar {\ensuremath{q\overline q}\xspace}
\def\u     {\ensuremath{u}\xspace}
\def\ubar  {\ensuremath{\overline u}\xspace}

\def\d     {\ensuremath{d}\xspace}

\def\c     {\ensuremath{c}\xspace}

\def\bbar  {\ensuremath{\overline b}\xspace}
\def\bbbar {\ensuremath{b\overline b}\xspace}

\def\piz   {\ensuremath{\pi^0}\xspace}

\def\pip   {\ensuremath{\pi^+}\xspace}
\def\pim   {\ensuremath{\pi^-}\xspace}

\def\pimp  {\ensuremath{\pi^\mp}\xspace}

\def\Kbar  {\kern 0.2em\overline{\kern -0.2em K}{}\xspace}

\def\Kz    {\ensuremath{K^0}\xspace}
\def\Kzb   {\ensuremath{\Kbar^0}\xspace}
\def\KzKzb {\ensuremath{\Kz \kern -0.16em \Kzb}\xspace}
\def\Kp    {\ensuremath{K^+}\xspace}
\def\Km    {\ensuremath{K^-}\xspace}

\def\KpKm  {\ensuremath{\Kp \kern -0.16em \Km}\xspace}

\def\Dbar    {\kern 0.2em\overline{\kern -0.2em D}{}\xspace}

\def\Dz      {\ensuremath{D^0}\xspace}
\def\Dzb     {\ensuremath{\Dbar^0}\xspace}
\def\DzDzb   {\ensuremath{\Dz {\kern -0.16em \Dzb}}\xspace}
\def\Dp      {\ensuremath{D^+}\xspace}
\def\Dm      {\ensuremath{D^-}\xspace}

\def\DpDm    {\ensuremath{\Dp {\kern -0.16em \Dm}}\xspace}

\def\B       {\ensuremath{B}\xspace}
\def\Bbar    {\kern 0.18em\overline{\kern -0.18em B}{}\xspace}

\def\BB      {\ensuremath{B\Bbar}\xspace} 
\def\Bz      {\ensuremath{B^0}\xspace}
\def\Bzb     {\ensuremath{\Bbar^0}\xspace}
\def\BzBzb   {\ensuremath{\Bz {\kern -0.16em \Bzb}}\xspace}
\def\Bu      {\ensuremath{B^+}\xspace}
\def\Bub     {\ensuremath{B^-}\xspace}
\def\Bp      {\ensuremath{\Bu}\xspace}

\def\Bpm     {\ensuremath{B^\pm}\xspace}

\def\BpBm    {\ensuremath{\Bu {\kern -0.16em \Bub}}\xspace}

\def\BorBbar    {\kern 0.18em\optbar{\kern -0.18em B}{}\xspace}
\def\DorDbar    {\kern 0.18em\optbar{\kern -0.18em D}{}\xspace}
\def\KorKbar    {\kern 0.18em\optbar{\kern -0.18em K}{}\xspace}

\mathchardef\Upsilon="7107
\def\Y#1S{\ensuremath{\Upsilon{(#1S)}}\xspace}% no space before {...}!

\def\FourS {\Y4S}

\mathchardef\Deltares="7101
\mathchardef\Xi="7104
\mathchardef\Lambda="7103
\mathchardef\Sigma="7106
\mathchardef\Omega="710A

\def\Deltabar{\kern 0.25em\overline{\kern -0.25em \Deltares}{}\xspace}
\def\Lbar{\kern 0.2em\overline{\kern -0.2em\Lambda\kern 0.05em}\kern-0.05em{}\xspace}
\def\Sigbar{\kern 0.2em\overline{\kern -0.2em \Sigma}{}\xspace}
\def\Xibar{\kern 0.2em\overline{\kern -0.2em \Xi}{}\xspace}
\def\Obar{\kern 0.2em\overline{\kern -0.2em \Omega}{}\xspace}
\def\Nbar{\kern 0.2em\overline{\kern -0.2em N}{}\xspace}
\def\Xb{\kern 0.2em\overline{\kern -0.2em X}{}\xspace}

\def\BR         {{\ensuremath{\cal B}\xspace}}

\def\pt         {\mbox{$p_T$}\xspace}
\def\mes        {\mbox{$m_{\rm ES}$}\xspace}

\def\DeltaE     {\mbox{$\Delta E$}\xspace}

\newcommand{\tev}{\ensuremath{\mathrm{\,Te\kern -0.1em V}}\xspace}
\newcommand{\gev}{\ensuremath{\mathrm{\,Ge\kern -0.1em V}}\xspace}
\newcommand{\mev}{\ensuremath{\mathrm{\,Me\kern -0.1em V}}\xspace}
\newcommand{\kev}{\ensuremath{\mathrm{\,ke\kern -0.1em V}}\xspace}
\newcommand{\ev}{\ensuremath{\mathrm{\,e\kern -0.1em V}}\xspace}
\newcommand{\gevc}{\ensuremath{{\mathrm{\,Ge\kern -0.1em V\!/}c}}\xspace}
\newcommand{\mevc}{\ensuremath{{\mathrm{\,Me\kern -0.1em V\!/}c}}\xspace}
\newcommand{\gevcc}{\ensuremath{{\mathrm{\,Ge\kern -0.1em V\!/}c^2}}\xspace}
\newcommand{\mevcc}{\ensuremath{{\mathrm{\,Me\kern -0.1em V\!/}c^2}}\xspace}
\def\invfb   {\ensuremath{\mbox{\,fb}^{-1}}\xspace}

\def\mus  {\ensuremath{\rm \,\mus}\xspace}

\def\ps   {\ensuremath{\rm \,ps}\xspace}

\def\mus        {\ensuremath{\,\mu{\rm s}}\xspace}    %% microsecond
\def\ps         {\ensuremath{{\rm \,ps}}\xspace}  %% picosecond

\def\calA{{\ensuremath{\cal A}}\xspace}

\def\to                 {\ensuremath{\rightarrow}\xspace}

\newcommand{\stat}{\ensuremath{\mathrm{(stat)}}\xspace}
\newcommand{\syst}{\ensuremath{\mathrm{(syst)}}\xspace}

\def\pep2{PEP-II}

\def\gsim{{~\raise.15em\hbox{$>$}\kern-.85em
          \lower.35em\hbox{$\sim$}~}\xspace}
\def\lsim{{~\raise.15em\hbox{$<$}\kern-.85em
          \lower.35em\hbox{$\sim$}~}\xspace}

\def\CP                {\ensuremath{C\!P}\xspace}
\def\C       {\ensuremath{C}\xspace}

\def\rhobar {\ensuremath{\overline \rho}\xspace}
\def\etabar {\ensuremath{\overline \eta}\xspace}
\def\mistag{\ensuremath{w}\xspace}

\def\deltaz{\ensuremath{{\rm \Delta}z}\xspace}
\def\deltat{\ensuremath{{\rm \Delta}t}\xspace}
\def\deltamd{\ensuremath{{\rm \Delta}m_d}\xspace}

\xspace

\def\jetset74   {\mbox{\tt Jetset \hspace{-0.5em}7.\hspace{-0.2em}4}\xspace}
\newcommand\vud {\ensuremath{V_{\mathrm{ud}}}}
\newcommand\vus {\ensuremath{V_{\mathrm{us}}}}
\newcommand\vub {\ensuremath{V_{\mathrm{ub}}}}
\newcommand\vcd {\ensuremath{V_{\mathrm{cd}}}}
\newcommand\vcs {\ensuremath{V_{\mathrm{cs}}}}
\newcommand\vcb {\ensuremath{V_{\mathrm{cb}}}}
\newcommand\vtd {\ensuremath{V_{\mathrm{td}}}}
\newcommand\vts {\ensuremath{V_{\mathrm{ts}}}}
\newcommand\vtb {\ensuremath{V_{\mathrm{tb}}}}
\def\vckm       {\ensuremath{{V}_{\rm CKM}}}
\def\theckmmatrix  {\ensuremath{ \left( \begin{array}{ccc} \vud & \vus & \vub \\ \vcd & \vcs & \vcb \\ \vtd & \vts & \vtb \end{array}\right).}}

\newcommand{\su}     [1]  {\ensuremath{SU(#1)}}

\newcommand{\e}      [1]   { {\ensuremath{ \times 10^{ {#1} } }}}
\def\br        {\ensuremath{ {\cal {B}}}}

\def\offpeak {off-peak}
\def\onpeak {on-peak}
\def\rhobar {\ensuremath{\overline{\rho}}}
\def\etabar {\ensuremath{\overline{\eta}}}

\def\Pdf    {\ensuremath{{\cal P}}}
\def\like   {\ensuremath{{\cal L}}}
\def\Bch   {\ensuremath{B^{\pm}}}

\def\mes   {\ensuremath{m_{ES}}}
\def\de   {\ensuremath{\Delta E}}
\def\dt   {\ensuremath{\Delta t}}

\def\coshel {\ensuremath{ cos( \theta_{H}) }}
\def\coshelone {\ensuremath{ cos( \theta_{H1}) }}
\def\cosheltwo {\ensuremath{ cos( \theta_{H2}) }}

\def\dm   {\ensuremath{\deltamd}}

\def\pt   {\ensuremath{ p_t }}

\def\rhoz   {\ensuremath{\rho^{0}}}

\def\qq   {\ensuremath{q \overline{q}}}
\def\bb   {\ensuremath{B \overline{B}}}
\def\ifb   {\ensuremath{\mbox{\,fb}^{-1}}\xspace}

\def\Brec  {\ensuremath{B_{\mathrm{rec}}}}
\def\Btag  {\ensuremath{B_{\mathrm{tag}}}}
\def\SCF   {{\ensuremath{\mathrm{SCF}}}}

\def\Long  {{\ensuremath{\mathrm{long}}}}
\def\Tran  {{\ensuremath{\mathrm{tran}}}}

\def\clong   {\ensuremath{C_{\Long}}}
\def\slong   {\ensuremath{S_{\Long}}}
\def\ctran   {\ensuremath{C_{\Tran}}}
\def\stran   {\ensuremath{S_{\Tran}}}

\def\ptrue   {\ensuremath{f_{L}}}
\def\nsig    {\ensuremath{N_{\rm signal}}}

\def\aeff {\ensuremath{\alpha_{\mathrm{eff}}}}

\def\Acp {\ensuremath{A_{\CP}}}
\def\piz {\ensuremath{\pi^0}}

\def\delC {\ensuremath{\Delta C}}
\def\delS {\ensuremath{\Delta S}}

\def\de       {\ensuremath{\Delta E}}

\def\borhorho {\ensuremath{B^{0} \rightarrow \rho^+ \rho^- }}
\def\brhorhoo {\ensuremath{B^{+} \rightarrow \rho^+ \rho^0 }}

\def\Bztorhoprhom {\ensuremath{\Bz (\Bzb) \to \rho^+ \rho^- }\xspace}

\def\rhop {\ensuremath{\rho^+ }\xspace}
\def\rhom {\ensuremath{\rho^- }\xspace}

\def\coshel  {\ensuremath{ \cos\theta_{i} }}
\def\mv      {\ensuremath{ m_{\pi^\pm \pi^0 }}}
\def\coshelone {\ensuremath{ \cos \theta_{1} }}
\def\mvone   {\ensuremath{ m_{1} }}
\def\cosheltwo {\ensuremath{ \cos \theta_{2} }}
\def\mvtwo   {\ensuremath{ m_{2} }}
\def\nno     {{\ensuremath{\cal{N}}}}

\def\Bztorhozrhoz {\ensuremath{\,\Bz \to \rho^0\rho^0}}
\def\Bztorhoprhom {\ensuremath{\,\Bz \to \rho^+\rho^-}}

\def\lepton   {{\sf Lepton}}

\def\kaonone  {{\sf Kaon I}}
\def\kaontwo  {{\sf Kaon II}}
\def\kaonpion {{\sf Kaon-Pion}}
\def\pion     {{\sf Pion}}
\def\other    {{\sf Other}}
\def\notag    {{\sf Untagged}}

\def\lumi {\ensuremath{349\, \ifb}}
\def\nbb {\ensuremath{(383.6\pm 4.2)\e{6}\, \bb\ \rm pairs}}
\def\offpeaklumi {\ensuremath{27.2\, \ifb}}
\def\ndata {37424} % number of fitted events
\def\extralumi{134.0}
\newcommand{\ImLambda}{\ensuremath{{\mathcal{I}m}\, \lambda_{\CP}}}

\def\correctedslong      {\ensuremath{-0.17 \pm 0.20 \stat}}
\def\correctedclong      {\ensuremath{0.01 \pm 0.15 \stat}}
\def\correctedfl         {\ensuremath{0.992 \pm 0.024 \stat}}
\def\correctedsignalyield{\ensuremath{729 \pm 60 \stat}}
\def\correctedbf         {\ensuremath{(25.5 \pm 2.1 \stat)\e{-6}}}
\def\sccorrelation       {\ensuremath{-0.035}}
\def\measuredslong {\ensuremath{\correctedslong ^{+0.05}_{-0.06} \syst}}
\def\measuredclong {\ensuremath{\correctedclong \pm 0.06 \syst}}
\def\measuredfl    {\ensuremath{\correctedfl ^{+0.026}_{-0.013} \syst}}
\def\measuredbf    {\ensuremath{(25.5 \pm 2.1 \stat ^{+3.6}_{-3.9} \syst)\e{-6}}}
\def\measuredalpha {\ensuremath{[73.1, 117.0]^\circ\, {\rm at}\, 68\%\, {\rm CL}}}

\def\measureddeltaalpha{\ensuremath{18^\circ}}
\def\deltatp{\ensuremath{\delta_{\mathrm{TP}}}}

\def\fitbiasonyield{\ensuremath{38.5\pm 5.3}}
\def\fitbiasonfl{\ensuremath{0.016\pm 0.002}}

\def\measuredalphamodel {\ensuremath{(89.8^{+7.0}_{-6.4})^\circ}}
\def\measuredalphamodelnodeltaconstraint {\ensuremath{[83.3,105.8]^\circ\, {\rm at}\, 68\%\, {\rm CL}}}

\def\modelboundonr{\ensuremath{0.10^{+0.03}_{-0.04}}}

\def\publishedmodelerroronalpha {\ensuremath{(+2, -5)^\circ}}

\def\utfitalpha{\ensuremath{(92.9 \pm 5.7)^\circ}}
\def\ckmfitalpha{\ensuremath{(100^{+4.5}_{-7.3})^\circ}}

\def\NonRes{\ensuremath{\rm NR}}
\def\classzero{\ensuremath{ \Bp  \to  K^+ \pi^+ \pi^-}}
\def\classone{\ensuremath{ \Bp  \to  \pi^+ \pi^0 \pi^0}}
\def\classtwo{\ensuremath{ \Bp  \to  \pi^0 \rho^+}}
\def\classthree{\ensuremath{ \Bp  \to  \rho^+ \rho^0}}
\def\classfour{\ensuremath{ \Bp  \to  \rho^+ \pi^+ \pi^- }}
\def\classfive{\ensuremath{ \Bp  \to  K_2^{*}(1770)  \rho}}
\def\classsix{\ensuremath{ \Bp  \to  K_0^{*} \pi }}
\def\classseven{\ensuremath{ \Bp  \to  a_1^+ \pi^0}}
\def\classeight{\ensuremath{ \Bp  \to  \rho^+ \pi^0 \pi^0}}
\def\classnine{\ensuremath{ \Bp  \to  a_{1}^0(\rho^+\pi^-) \rho^+}}
\def\classten{\ensuremath{ \Bp  \to  a_1^0 \pi^+}}

\def\classtwelve{\ensuremath{ \Bp \to \mathrm{charmless}}}

\def\classthirteen{\ensuremath{ \Bz  \to  \pi^+ \pi^- \pi^0\,  (\NonRes)}}
\def\classfourteen{\ensuremath{ \Bz  \to  \rho^\pm \pi^\mp}}
\def\classfifteen{\ensuremath{ \Bz  \to  \piz\piz K^0_S}}
\def\classsixteen{\ensuremath{ \Bz  \to  a_{1}^\pm(\rho^0 \pi^\pm) \pi^\mp}}
\def\classseventeen{\ensuremath{ \Bz  \to  \rho^\mp \pi^\pm \pi^0}}
\def\classeighteen{\ensuremath{ \Bz  \to  K_2^*(1770) \rho}}
\def\classnineteen{\ensuremath{ \Bz \to a_{1}^\pm(\rho^\pm \pi^0) \rho^\mp}}
\def\classtwenty{\ensuremath{ \Bz \to a_{1}^\pm(\rho^0 \pi ^\pm) \rho^\mp}}
\def\classtwentyone{\ensuremath{ \Bz  \to  a_1^\pm(\rho^\pm \pi^0) \pi^\mp}}

\def\classtwentythree{\ensuremath{ \Bz \to \mathrm{charmless}}}

\def\FF {\ensuremath{FF}}
\def\FT {\ensuremath{FT}}
\def\TF {\ensuremath{TF}}
\def\TT {\ensuremath{TT}}

\newcommand{\SLACPubNumber} {12498}

\def\figurebox#1#2#3{%
    \def\arg{#3}%
    \ifx\arg\empty
    {\hfill\vbox{\hsize#2\hrule\hbox to #2{\vrule\hfill\vbox to #1{\hsize#2\vfill}\vrule}\hrule}\hfill}%
    \else
    {\hfill\epsfbox{#3}\hfill}%
    \fi}

\newcommand {\vecp} {\ensuremath{\kern 0.2em\vec{\kern 0.1em p}{}\xspace}}
\newcounter{Lcount}

\begin{document}

%\preprint{\babar-PUB-\BABARPubYear/\BABARPubNumber} 
%\preprint{SLAC-PUB-\SLACPubNumber} 

\begin{flushleft}
%BAD1612 Version 9\\
%\babar-PUB-\BABARPubYear/\BABARPubNumber\\ 
SLAC-PUB-\SLACPubNumber\\[10mm]
%hep-ex/\LANLNumber\\[10mm]
\end{flushleft}

\title{
{\large \bf A Study of {\boldmath \Bztorhoprhom} Decays and Constraints on the CKM Angle {\boldmath $\alpha$}.}}

% Dummy author list; contact PubBoard Chair for current author list
%\input ../pubboard/authors_feb2007_bad1612.tex
%% author list as of 02-Feb-2007 (578 authors)
%
\author{B.~Aubert}
\author{M.~Bona}
\author{D.~Boutigny}
\author{Y.~Karyotakis}
\author{J.~P.~Lees}
\author{V.~Poireau}
\author{X.~Prudent}
\author{V.~Tisserand}
\author{A.~Zghiche}
\affiliation{Laboratoire de Physique des Particules, IN2P3/CNRS et Universit\'e de Savoie, F-74941 Annecy-Le-Vieux, France }
\author{J.~Garra~Tico}
\author{E.~Grauges}
\affiliation{Universitat de Barcelona, Facultat de Fisica, Departament ECM, E-08028 Barcelona, Spain }
% ...
\collaboration{The \babar\ Collaboration}
\noaffiliation

%\date{\today}% It is always \today, today, but you may specify any date with \date.

% Abstract
%----------------
\begin{abstract}
We present results from an analysis of \Bztorhoprhom\ decays using
\nbb\ collected by the \babar\ detector at the
\pep2\ asymmetric-energy $B$ Factory at SLAC.  The measurements of the
\Bztorhoprhom\ branching fraction, longitudinal polarization fraction 
\ptrue, and the \CP-violating parameters \slong\ and \clong\ are:
\begin{eqnarray}
{\cal B}(\Bztorhoprhom) &=& \measuredbf,\nonumber \\
\ptrue &=& \measuredfl,\nonumber \\
\slong &=& \measuredslong,\nonumber \\
\clong &=& \measuredclong.\nonumber
\end{eqnarray}
We determine the unitarity triangle angle $\alpha$, using an isospin analysis of $B\rightarrow \rho\rho$ decays. 
One of the two solutions, $\alpha = \measuredalpha$ is compatible with 
standard model-based fits of existing data.
Constraints on the unitarity triangle are also evaluated using an \su{3} symmetry based approach.
\end{abstract}

% PACS, the Physics and Astronomy Classification Scheme.

\maketitle

%\input{introduction}
%-----------------------
% Introduction
%-----------------------
\section{\boldmath Introduction\label{sec:introduction}}

\begin{figure}[h!]
\begin{center}
  \resizebox{6cm}{!}{\includegraphics{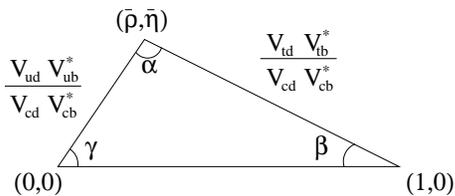}}
\end{center}
 \caption{The Unitarity Triangle in the $\rhobar$-$\etabar$ plane.}
\label{fig:unitarity_triangle}
\end{figure}

Charge conjugation-parity (\CP) violation was first seen in the decay of neutral kaons~\cite{christenson}.
It was shown some forty years ago that \CP violation is a necessary but insufficient condition
required to generate a net baryon anti-baryon asymmetry in the universe~\cite{asakharov}.
The standard model (SM) of electroweak interactions describes \CP\ violation
as a consequence of a complex phase in the three-generation Cabibbo-Kobayashi-Maskawa (CKM) 
quark-mixing matrix~\cite{CKM1,CKM2}:
\begin{eqnarray}
\vckm = \theckmmatrix
\end{eqnarray}
\vckm\ describes the couplings of the $u$, $c$ and $t$ quarks to
$d$, $s$ and $b$ quarks, which are mediated by the exchange of a \W\ boson.  
In \B-meson decays the \CP\ violating parameters of the SM are most directly
related to the angles and sides of the so-called Unitarity Triangle (UT), shown in 
Fig.~\ref{fig:unitarity_triangle}. 
The angles $\alpha$, $\beta$, and $\gamma$ are defined as
\begin{eqnarray}
\alpha \equiv \arg\left[-\vtd\vtb^*/\vud\vub^*\right], \\
\beta  \equiv \arg\left[ -\vcd\vcb^* / \vtd\vtb^*\right], \\
\gamma \equiv \arg\left[ -\vud\vub^* / \vcd\vcb^* \right].
\end{eqnarray}
Any non-trivial phase in $V_{\mathrm{ij}}$ is \CP violating.  
\CP violating phases originating from the CKM matrix are related
to weak interactions, and therefore referred to as weak phases.
In the Wolfenstein convention~\cite{wolfenstein}, the angle $\gamma$ is the phase of \vub, $\beta$ is the phase of \vtd, and $\alpha$ is 
the phase difference between \vub\ and \vtd\ constrained to satisfy $\alpha=\pi-\beta-\gamma$ through the unitarity of $\vckm$.
%----------------------------------
% Figure : The Unitarity Triangle.
%----------------------------------
%\begin{figure}[!ht]
%----------------------------------

In Fig.~\ref{fig:unitarity_triangle} the UT is shown in the complex $(\rhobar, \etabar)$ plane, 
where the apex is given by the phase-convention independent definition~\cite{buras}:
\begin{equation}
\rhobar + i\etabar \equiv -\frac{\vud\vub^*}{\vcd\vcb^*}.
\end{equation}

The quest to understand \CP violation remains, despite its observation in the $B$ meson
system by both the \babar~\cite{babar_sin2beta_2002} and Belle experiments~\cite{belle_sin2beta_2002},
since the SM does not, through the CKM phase, incorporate enough \CP\ violation to explain the 
current matter-antimatter asymmetry~\cite{gavela,huet}.     
The CKM angle $\beta$ is measured to a 1$^{\circ}$ precision with 
$b\to c\overline{c}s$ decays~\cite{babar_sin2beta,belle_sin2beta} and is consistent with current 
predictions~\cite{ref:ciuchini1995,ckmfitter,utfitter}.
A significant deviation from results of SM-based fits of existing data for any of the UT angles would be a clear indication of new physics~\cite{bigiandsanda,brancolavouraandsilva}.

\begin{figure}[!ht]
\begin{center}
\resizebox{5cm}{!}{\includegraphics{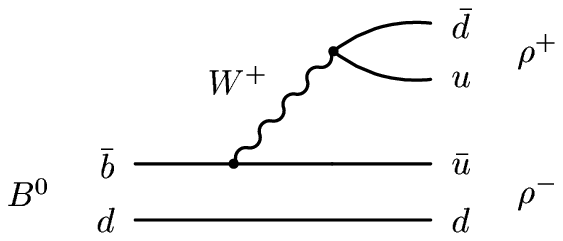}}
\resizebox{5cm}{!}{\includegraphics{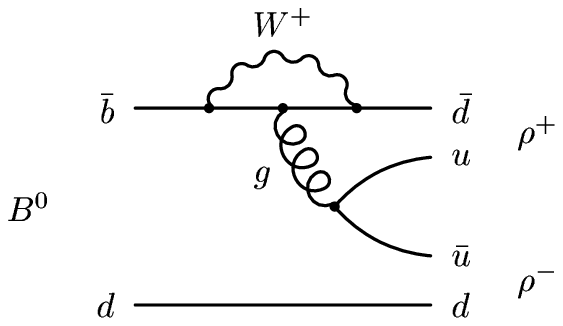}}
\caption{Tree (top) and gluonic penguin (bottom) diagrams contributing to the process \borhorho. The
         penguin contribution coming from the diagram with a top quark in the loop dominates
         as contributions from processes with \u\ and \c\ quarks are suppressed.}
\label{fig:feynmangraphs}
\end{center}
\end{figure}

In principle, direct experimental measurements of $\alpha$ can be made from decays that proceed mainly 
through a $\bbar \to \u\ubar \d$ tree diagram such as $\Bztorhoprhom, \rho^\pm\pi^\mp, \pi^+\pi^-$
and $a_1\pi$ ~\cite{conj,bevan2006}.
Interference between the direct decay and decay after $\Bz\Bzb$ mixing in each of these decays results 
in a time-dependent decay-rate asymmetry that is sensitive to the angle $\alpha$.
Figure~\ref{fig:feynmangraphs} shows the leading order tree and gluonic penguin loop contributions to the decay \Bztorhoprhom. 
The presence of penguin contributions complicates the extraction of $\alpha$ from these decays.  
Using isospin relations~\cite{gronaulondon}, measurements of the $\Bp \to \rho^+\rho^0$~\cite{babarrhoprhoz,bellerhorho0} and
$\Bz\to \rho^0 \rho^0$~\cite{babarrhozrhoz} branching fractions show that the penguin
contribution in $\B \to \rho \rho$ is smaller than the leading tree diagram.
The use of \su{3} flavor symmetry to increase the precision on the weak phase constrained using $\Bz\to\rho^+\rho^-$ and $B^+\to K^{*0}\rho^{+}$ decays 
has also been proposed~\cite{ref:benekesuthree}.  Both of these approaches are discussed in Section~\ref{sec:alpha}.  
Section~\ref{sec:alpha} and Ref.~\cite{ref:zupanckm} describe a 
number of possible sources of theoretical uncertainty.

%-------------------------------------
% Figure : Tree and penguin diagrams
%-------------------------------------
%----------------------------------------------------------------------------------------------

\begin{figure}[!ht]
\begin{center}
\resizebox{6.5cm}{!}{\includegraphics{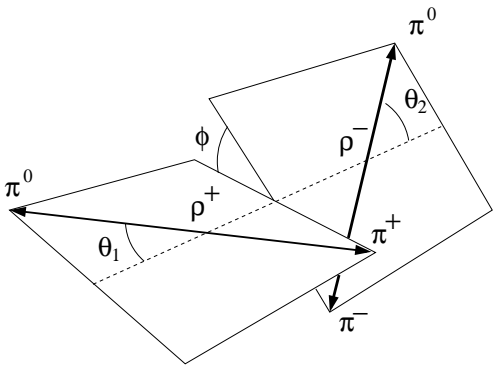}}
\caption{A schematic of the decay of a \B\ meson via two $\rho$ mesons to a four pion final state.
  The $\rho$ meson final states are shown in their rest frames, and $\phi$ is the angle
  between the decay planes of the $\rho$ mesons.}
\label{fig:btovv}
\end{center}
\end{figure}

%
% B->VV
%
In $\Bz\to\rho^+\rho^-$ decays, a spin 0 particle (the \Bz\ meson) decays into two spin 1 
particles ($\rho^\pm$ mesons), as shown in Fig.~\ref{fig:btovv}.  Subsequently each $\rho^\pm$ meson decays into 
a $\pi^\pm\pi^0$ pair.  As a result, the \CP analysis of $\Bz\to \rho^+ \rho^-$ is complicated 
by the presence of one amplitude with longitudinal polarization and 
two amplitudes with transverse polarization.  The longitudinal amplitude is \CP-even, 
while the transverse amplitudes contain \CP-even and \CP-odd states.
The decay is observed to be dominated by the longitudinal 
polarization \cite{babarrhoprhomr14,bellerhoprhom}, with a fraction $\ptrue$ defined as the 
fraction of the helicity zero state in the decay. Integrating over the angle between the 
$\rho$ decay planes $\phi$, the angular decay rate is
\begin{eqnarray}
\label{eqn:one} 
\frac{d^2\Gamma}{\Gamma d\cos\theta_1 d\cos\theta_2}= \frac{9}{4} \Bigg{[}f_L \cos^2\theta_1 \cos^2\theta_2 \nonumber \\
 + \frac{1}{4}(1-\ptrue) \sin^2\theta_1 \sin^2\theta_2 \Bigg{]}, 
\end{eqnarray}
where the helicity angles $\theta_{i=1,2}$ are the angles between the \piz\ momentum 
and the direction opposite to that of the $B^0$ in the $\rho$ rest 
frame.

%-------------------------------------
% Figure : B -> VV
%-------------------------------------
%----------------------------------------------------------------------------------------------

We identify (tag) the initial flavor of the reconstructed \B\ candidate,
\Brec, using information from the other \B\ meson, \Btag, in the event.
The decay rate $f_+(f_-)$ for a neutral \B\ meson decaying into a \CP\
eigenstate, where the \Btag\ is a \Bz $(\Bzb)$, is given by:
\begin{eqnarray}
f_{\pm}(\deltat) &=& \frac{e^{ -\left|\deltat\right| / \tau_{\Bz}}}{4\tau_{\Bz}} \bigg{\{}1 \nonumber \\
   &&  \pm \eta_f [S\sin(\deltamd\deltat) - \C\cos(\deltamd\deltat)]\bigg{\}}.\nonumber \\
\label{eq:deltatdistribution}
\end{eqnarray}
Here \deltat\ is the proper time difference between the decay of the
  \Brec\ and \Btag\ mesons in an event, $\tau_{\Bz}$ 
is the mean \Bz\ lifetime, $\deltamd$ is the \Bz-\Bzb\ mixing frequency~\cite{pdg2006},
and $\eta_f$ is the \CP\ eigenvalue of the decay. 
This assumes that there is no difference between the \Bz\ and \Bzb\ lifetimes, $\Delta \Gamma = 0$. 
The parameters $S$ and $C$ describe \B-mixing induced and direct \CP\ violation, respectively.
For the longitudinal polarization $\eta_f=+1$, $S=\slong$ and $C=\clong$, are defined as 
\begin{eqnarray}
\slong = \frac{ 2\, \ImLambda }{ 1 + |\lambda_{\CP}|^2}, \label{eq:slong}\\
\clong = \frac{ 1 - |\lambda_{\CP}|^2 }{ 1 + |\lambda_{\CP}|^2},\label{eq:clong}
\end{eqnarray}
where $\lambda_{\CP}=\frac{q}{p}\frac{\overline{A}}{A}$~\cite{ref:lambda}, 
$q$ and $p$ are complex constants that relate the \B\ meson flavor eigenstates to the
mass eigenstates, and $\overline{A}/A$ is the ratio of amplitudes
of the decay of a \Bzb\ or \Bz\ to the final state under study.
\CP\ violation is probed by studying the time-dependent decay-rate asymmetry
\begin{equation}
 {\cal A}(\deltat) = \frac{
  \Gamma (\deltat) - \overline{\Gamma}(\deltat) } { \Gamma (\deltat) + \overline{\Gamma}(\deltat) },\label{eq:timedepasym}
\end{equation}
where $\Gamma$($\overline{\Gamma}$) is the decay-rate for \Bz(\Bzb) tagged
events.  This asymmetry has the form
\begin{equation}
{\calA}(\deltat)=S\sin(\deltamd\deltat)-\C\cos(\deltamd\deltat).
\end{equation}
The transverse polarization is an admixture of \CP-even ($\eta_f = +1$) and \CP-odd ($\eta_f = -1$) parts, 
where each \CP\ eigenstate has a decay-rate distribution of the form of 
Eq. (\ref{eq:deltatdistribution}). As the longitudinal polarization dominates 
this decay, we describe the \CP\ admixture of the transverse polarization 
with common parameters \stran\ and \ctran. We set $\stran=\ctran=0$, and 
vary these parameters when evaluating systematic uncertainties.

In the absence of penguin contributions in $\Bz \to \rho^+\rho^-$, 
$\slong = \sin 2 \alpha$ and $\clong = 0$.
The presence of penguin contributions with different weak phases
to the tree level amplitude shifts the experimentally measurable
parameter \aeff\ away from the value of $\alpha$. 
In the presence of penguin contributions $\aeff = \alpha +
\delta\alpha$, where 
\begin{equation}
\slong = \sqrt{1-\clong^2}\sin 2 \aeff,
\label{eq:seff}
\end{equation}
 and \clong\ can be non-zero.

For $\brhorhoo$ one measures a time-integrated charge asymmetry, which is analogous to Eq. (\ref{eq:timedepasym}) and 
is defined as 
\begin{eqnarray}
  \Acp &=& \frac{\overline{N}-N}{\overline{N}+N},
\end{eqnarray} 
where $N$ ($\overline{N}$) is the number of $B$ ($\overline{B}$) decays to this final state. 
Table~\ref{tbl:rhorhodataexistingresults} summarizes the most recent measurements of the 
complete set of experimental inputs needed to extract $\alpha$ from the $\B\to\rho\rho$ decays.

%------------------------------------------------------------------------------------------------------------
% Table : Past rho rho results
%------------------------------------------------------------------------------------------------------------
\begin{table*}[!t]
\caption{Measurements of the $B \to \rho\rho$ decays.  Branching
         fractions (\br) are in units of $10^{-6}$. The column marked \clong (\Acp) denotes
         the time dependent charge asymmetry \clong\ in the case of the \Bztorhoprhom\
         decay, and the time integrated charge asymmetry \Acp\ in the case of
         $\Bch \to \rho^\pm\rho^0$.}
\small
\begin{center}
\begin{tabular}{c|ccccc}\hline\hline\vspace{-0.3cm}\\
Mode                      & Expt. (luminosity \ifb) & \br\ (\e{-6}) & \ptrue & \clong (\Acp) & \slong \\[3pt] \hline\noalign{\vskip1pt}
\Bztorhoprhom             & \babar~\cite{babarrhoprhomprlr12} (82)& $30 \pm 4 \pm 5$ & $-$ & $-$ & $-$ \\
\Bztorhoprhom             & \babar~\cite{babarrhoprhomr14} (210) & $-$                        & $0.978\pm 0.014 ^{+0.021}_{-0.029}$  & $\ \ -0.03 \pm 0.18 \pm 0.09$ &  $\ \ -0.33 \pm 0.24^{+0.08}_{-0.14}$\\
\Bztorhoprhom             & Belle~\cite{bellerhoprhom} (253)& $22.8 \pm 3.8 ^{+2.3}_{-2.6}$ & $0.941^{+0.034}_{-0.040} \pm 0.030$ & 
                            $-$ & $-$ \\
\Bztorhoprhom             & Belle~\cite{bellerhoprhomupdate} (535) & $-$ & $-$ & $-0.16 \pm 0.21 \pm 0.08$\footnote{Belle Collaboration uses the opposite sign convention for \clong.} & $0.19 \pm 0.30 \pm 0.08$ \\
$\Bch \to \rho^\pm\rho^0$ & \babar~\cite{babarrhoprhoz} (210)  & $16.8 \pm 2.2 \pm 2.3$       & $0.905 \pm 0.042^{+0.023}_{-0.027}$  & $-0.12 \pm 0.13 \pm 0.10$ & $-$ \\
$\Bch \to \rho^\pm\rho^0$ & Belle~\cite{bellerhorho0} (78)    & $31.7 \pm 7.1^{+3.8}_{-6.7}$ & $0.95 \pm 0.11 \pm 0.02$  & $0.00 \pm 0.22 \pm 0.03$ & $-$ \\
$\Bz \to \rho^0\rho^0$    & \babar~\cite{babarrhozrhoz} (349) & $1.07 \pm 0.33 \pm 0.19$     & $0.87 \pm 0.13 \pm 0.04$                      & $-$
           & $-$ \\[6pt] \hline\hline
\end{tabular}
\end{center}
\label{tbl:rhorhodataexistingresults}
\end{table*}
%----------------------------------------------------------------------------------------

In this article, we present an update of the time-dependent analysis of \Bztorhoprhom\ and measurement of the CKM angle $\alpha$ reported
in Ref.~\cite{babarrhoprhomr14} and branching fraction reported in Ref.~\cite{babarrhoprhomprlr12}.  Improvements to the analysis reported
here include an increase in data sample analyzed, a tighter constraint on the proper time difference between the two \B\ meson decays
in selected events, an improved algorithm to determine the flavor of \B\ mesons, a modified multivariate analyzer for background
suppression, and an improved description of the signal and background model.

%\input{detector}
%--------------------------------------------------------------
% Dataset and detector 
\section{\boldmath The Dataset and \babar\ Detector\label{sec:dataset}}
%--------------------------------------------------------------
The results presented in this paper are based on data collected 
with the \babar\ detector at the \pep2\ asymmetric \epem\ storage ring~\cite{ref:pepcdr}
 operating at the Stanford Linear Accelerator Center. At \pep2, 
9.0 \gev\ electrons and 3.1 \gev\ positrons are collided at a center-of-mass
energy of 10.58 \gev which corresponds to the mass of the \FourS\ resonance.  
The asymmetric energies result in a boost from the laboratory to the 
center-of-mass (CM) frame of $\beta\gamma\approx 0.56$.
The dataset analyzed has an integrated luminosity of \lumi\
corresponding to \nbb\ recorded at the \FourS\ resonance (\onpeak). 
An additional \offpeaklumi\ of data were recorded about 40 \mev\ below this energy 
(\offpeak) for the study of continuum background,
where light quarks are produced in the process $\epem\to\qqbar$ ($q = u,d,s,c$).

The \babar\ detector is described in detail elsewhere~\cite{babar_nim}.
Surrounding the interaction point is a five double-sided layer 
silicon vertex tracker (SVT) which measures the impact parameters of 
charged particle tracks in both the plane transverse to, and along 
the beam direction. A 40-layer drift chamber (DCH) surrounds the SVT 
and provides measurements of the momenta for charged 
particles. 
Both the SVT and DCH are surrounded by a solenoid magnet, that provides
a 1.5 T magnetic field.
Charged hadron identification is achieved 
through measurements of particle energy-loss in the tracking system 
and the Cherenkov angle obtained from a detector of internally 
reflected Cherenkov light. A CsI(Tl) electromagnetic calorimeter 
(EMC) provides photon detection, electron identification, and 
$\piz$ reconstruction. Finally, the instrumented flux return of 
the magnet allows discrimination of muons from pions. 
For the most recent $\extralumi\invfb$ of data, a portion of the 
resistive plate chambers constituting the muon system 
has been replaced by limited streamer tubes~\cite{ref:lsta,ref:lstb,ref:lstc}.

We use a right-handed coordinate system with the $z$ axis along the electron beam
direction and the $y$ axis upward, with the origin at the nominal beam interaction
point. Unless otherwise stated, kinematic quantities are
calculated in the laboratory rest frame.
The other reference frame which we commonly use is the CM frame 
of the colliding electrons and positrons.

We use Monte Carlo (MC) simulated events generated using the GEANT4~\cite{ref:geant} 
based \babar\ simulation.

%\input{reconstruction}
%-----------------------
% B reconstruction
%-----------------------
\section{\boldmath Reconstruction of \B\ candidates\label{sec:reconstruction}}
%---------------------

\subsection{\boldmath Photon and \piz\ reconstruction}

Photons are reconstructed from localized energy deposits in the EMC
that are not associated with a charged track.  We require 
photon candidates to have an energy greater than 50\mev, and a lateral
shower profile~\cite{ref:lat} to be consistent with the photon hypothesis.
We reconstruct neutral pions from combinations of two
distinct photon candidates where the invariant $\gamma\gamma$ mass $m_{\gamma\gamma}$ satisfies
$0.10 < m_{\gamma\gamma} < 0.16$ \gevcc.

\subsection{\boldmath $\rho^\pm$ reconstruction}
\label{sec:rhomasshelicity}

We combine reconstructed \piz\ mesons with charged tracks that are
consistent with the $\pi^\pm$ hypothesis to form $\rho^\pm$ candidates.
The invariant mass $m_{\pi^\pm\pi^0}$ of the reconstructed 
$\rho^\pm$ is required to lie between 0.5 and 1.0 \gevcc, to select events
in the vicinity of the $\rho$ resonance.  We require that the helicity angle
of each $\rho$ meson satisfies $-0.90 < \cos\theta_i < 0.98$.  This
selection criteria suppresses continuum and \B\ backgrounds.

%-----------------------------------------------------------------
\subsection{\boldmath \Bz\ reconstruction and event selection}
\label{sec:recoevsel}
%-----------------------------------------------------------------
We combine two oppositely charged $\rho$ candidates to form the \B\ candidate 
$\Brec$.  All particles in the rest of the event (ROE) are combined to 
form the other \B\ meson in the event \Btag. In addition to the two 
charged tracks in the $\Brec$, we require that there is at least one 
track in the \Btag.

In order to suppress potential backgrounds from $\epem \to e^+e^-$, $\mu^+\mu^-$ events, 
we require the second-to-zeroth Fox-Wolfram moment $R_2$~\cite{foxwolfram} of 
the event to be less than 0.98. 
Continuum events are the dominant background which is reduced by requiring 
the absolute value of the cosine of the angle between the $B_{rec}$ 
thrust axis ($TB$) and that of the ROE ($TR$) to satisfy $|\cos(TB, TR)|< 0.8$.  
We retain 17.1\% and 20.1\% of longitudinal and transverse signal, respectively, 
by requiring the aforementioned selection criteria.

We calculate $\deltat = \deltaz/\beta\gamma c$
from the measured separation \deltaz\ between the \Brec\ and \Btag\
vertices~\cite{babarsin2betaprd}. We determine the \Brec\ vertex
from the two charged-pion tracks in its decay. The \Btag\
decay vertex is obtained by fitting the other tracks in the event,
with constraints from the \Brec\ momentum and the
beam-spot location. The RMS resolution on
$\deltat$ is 1.1~\ps. We only use events that satisfy 
$|\deltat|<15~\ps$ and for which the error on \deltat ($\sigma_{\dt}$)
is less than $2.5~\ps$.

Signal events are identified using two kinematic variables, the
difference \DeltaE between the CM energy of the $B_{rec}$, $E_B^*$, and
$\sqrt{s}/2$, 
\begin{equation}
\DeltaE = E_B^* - \sqrt{s}/2,
\end{equation}
and the beam-energy-substituted mass,
\begin{equation}
\mes = \sqrt{(s/2 + {\mathbf {p}}_i\cdot {\mathbf {p}}_B)^2/E_i^2- {\mathbf {p}}_B^2},
\end{equation}
where $\sqrt{s}$ is the total CM energy. The  $B_{rec}$ momentum
${\mathbf {p}_B}$ and four-momentum of the initial state $(E_i, {\mathbf
{p}_i})$ are defined in the laboratory frame. We accept candidates that
satisfy $5.25 < \mes <5.29~\gevcc$ and $-0.12<\DeltaE<0.15~\gev$.
An asymmetric \DeltaE\ selection is used in order to reduce backgrounds 
from higher-multiplicity \B decays.  The resolution on \mes\ is 
dominated by the knowledge of the energy of the $e^+$ and $e^-$ beams,
and the resolution on \DeltaE\ is dominated by the reconstruction 
performance of the EMC.

After the selection criteria mentioned above have been applied, the
average number of candidates per event is approximately 1.5.
In events with more than one candidate, we select the candidate that minimizes the sum of 
$( m_{\gamma\gamma} - m_{\piz} )^2$ where $m_{\piz}$ is the true 
\piz\ mass~\cite{pdg2006}. In 0.3\% of events, the same \piz\ mesons are 
used by multiple \B\ candidates. In such an event we randomly select the 
candidate to keep.

%\input{backgroundsuppression}
%-----------------------
% Background suppression
%-----------------------
\section{\boldmath Continuum background suppression\label{sec:backgroundsupression}}

\begin{figure*}[!ht]
\begin{center}
 \resizebox{15cm}{!}{ \includegraphics{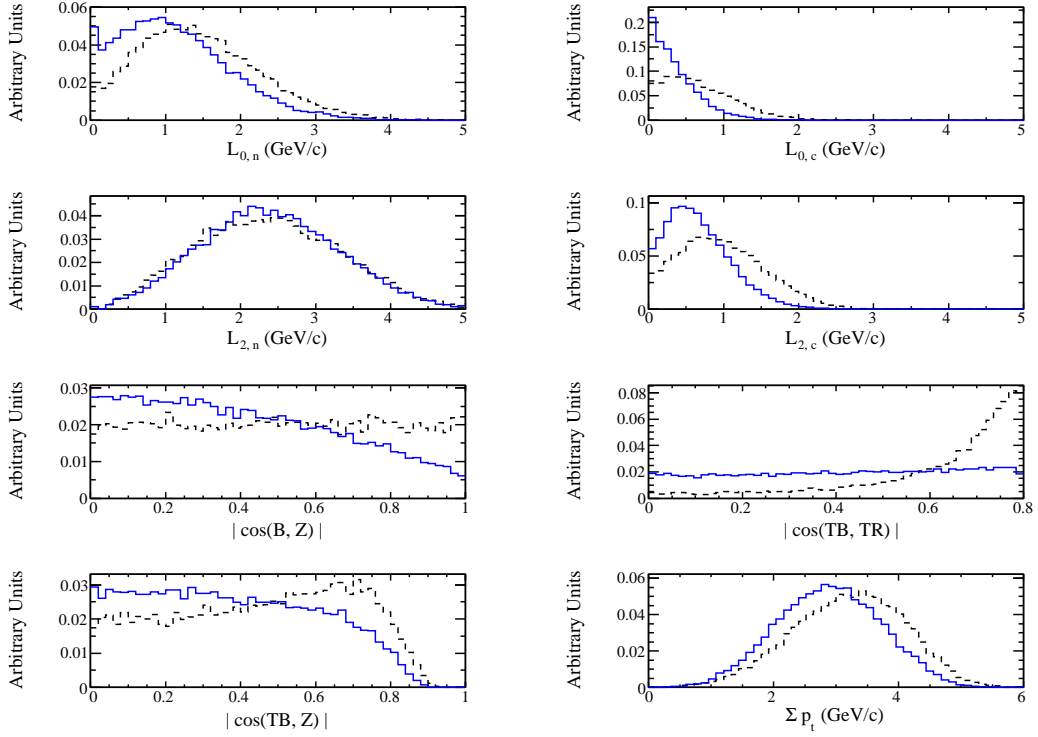} }
\caption{The input variables used in training the neural network.  The solid line 
         represents the signal training sample (MC simulated events) and the dashed 
	 line represents the continuum background (off-peak data).  The distributions shown are
         (in order left to right; top to bottom) $L_{0,n}, L_{0,c}$, $L_{2,n}, L_{2,c}$, 
         $\cos(B, Z)$, $\cos(B, TR)$, $\cos(TB, Z)$, and the sum of the \pt\ in the ROE.}
\label{fig:nnoinputvars}
\end{center}
\end{figure*}

\begin{figure*}[!ht]
\begin{center}
\resizebox{15cm}{!}{
 \includegraphics{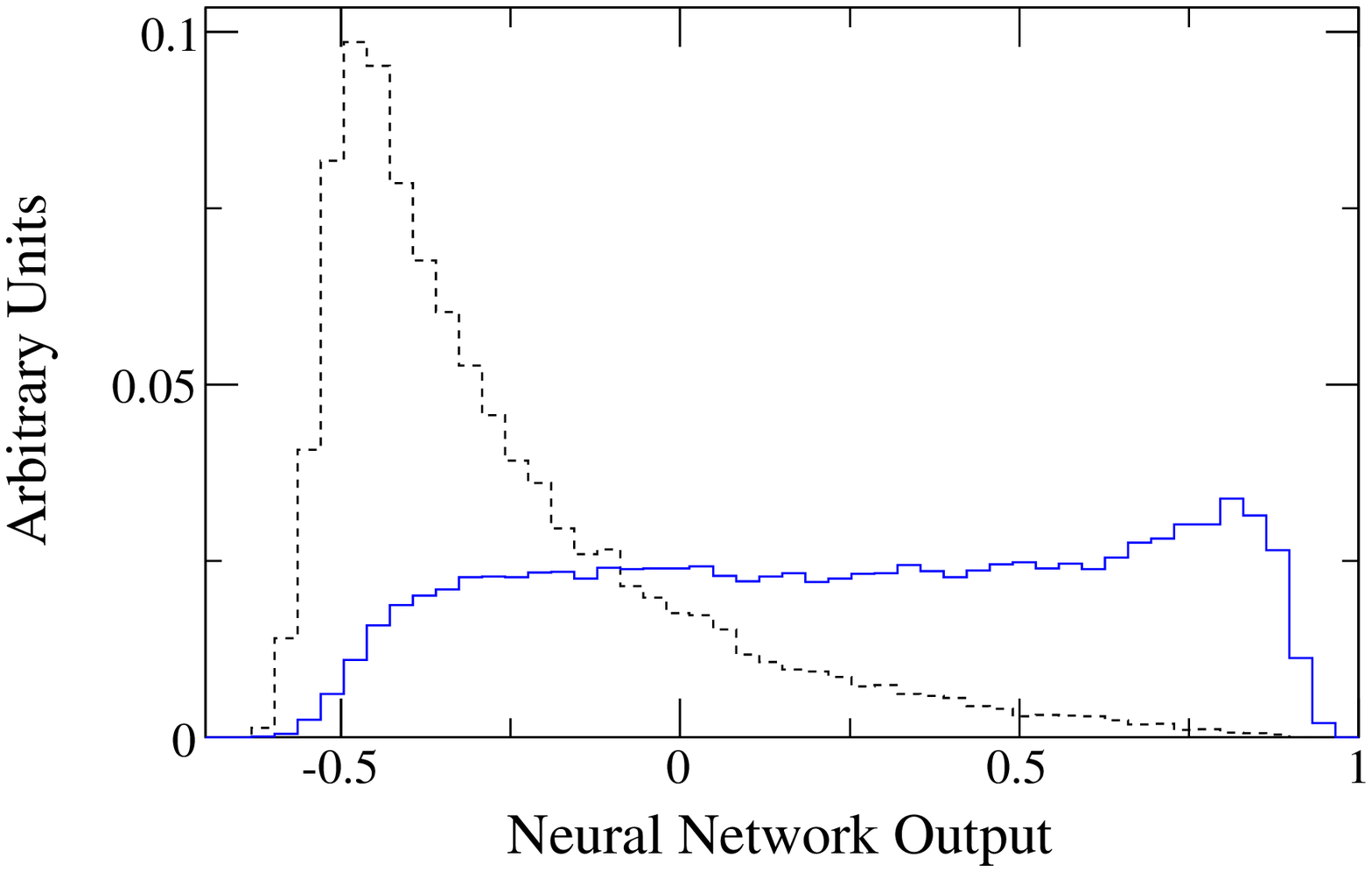}
 \includegraphics{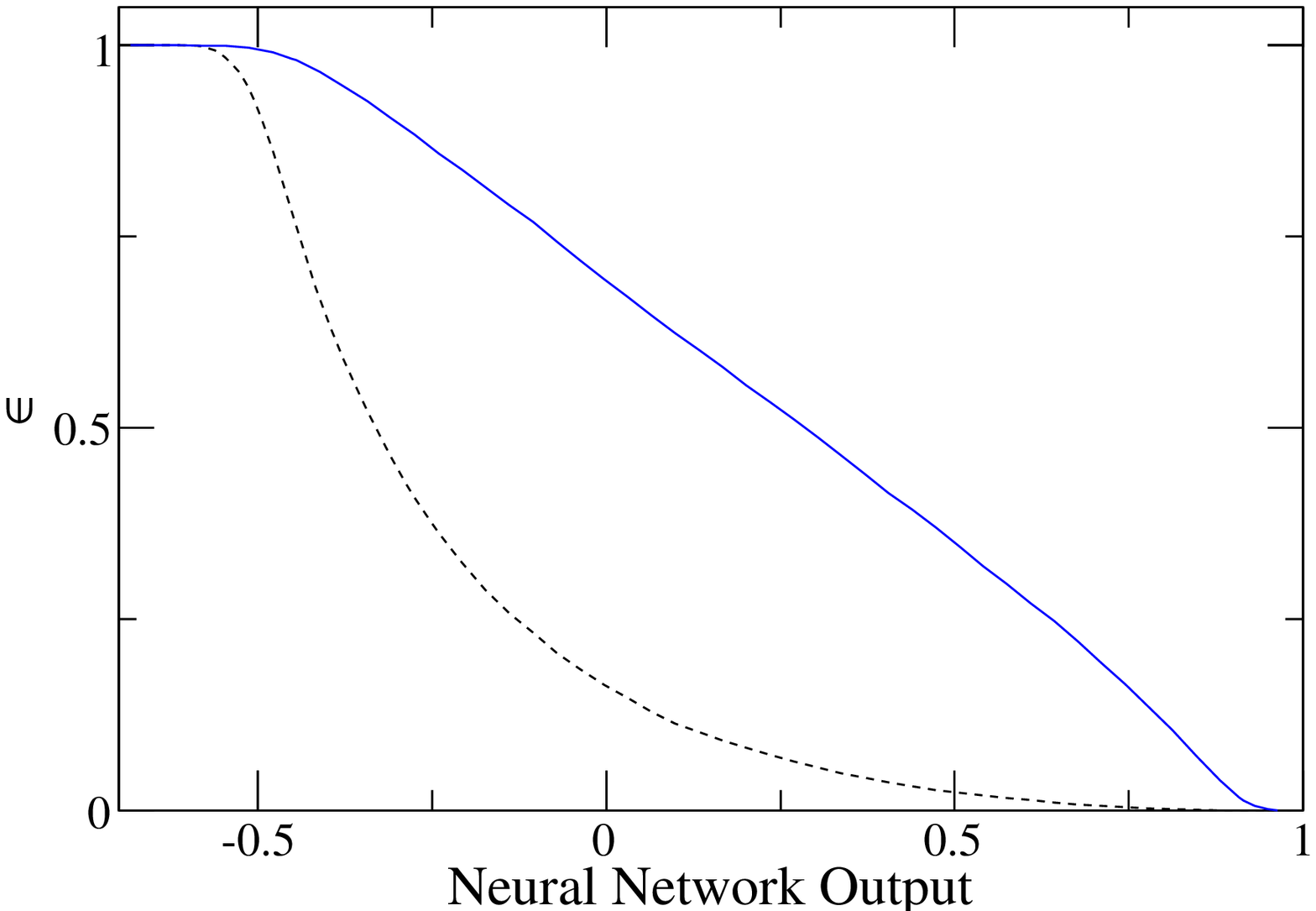}
}
\caption{The left plot shows the distribution of $\nno$ for off-peak data (dashed) and longitudinally 
polarized signal MC simulated events (solid). 
         The right plot shows the signal efficiency (solid) and continuum background efficiency (dashed) distribution as a function of cut on the minimum value of $\nno$.}
\label{fig:nnotraining}
\end{center}
\end{figure*}

In addition to the cuts on $\cos\theta_{i}$, $R_2$, and $|\cos(TB, TR)|$ that directly remove 
background events, we use an artificial neural network in order to 
discriminate between continuum background and signal events. For this 
purpose we combine the following eight variables into a single output, $\nno$.
\begin{itemize}
  \item The coefficients, $L_0, L_2$, split into sums 
    over the ROE for neutral and charged particles; 
    $L_{0,n}, L_{2,n}$ and $L_{0,c}, L_{2,c}$. The coefficients are defined
    as $L_k = \sum p_j |\cos(\psi_j)|^k$,
    where $k=0,2$, $p_j$ is the particle 
    momentum and $\psi_j$ is the angle of the particle direction 
    relative to the thrust axis of the \B\-candidate. Both $p_j$ and
    $\psi_j$ are defined in the CM frame.

  \item $|\cos(B, Z)|$, the absolute value of the cosine of the 
  angle between the direction of the \B\ and $z$ axis in the CM frame.  
  This variable follows a sine squared distribution for \bb\ events, whereas it 
  is almost uniform for \qq.

  \item $|\cos(TB, TR)|$. This variable, previously defined in Section~\ref{sec:recoevsel}, is strongly 
  peaked at unity for \qq\ events.  \bb\ events are more isotropic as 
  the \B\ mesons are produced close to the kinematic threshold.

  \item $|\cos(TB, Z)|$, the absolute value of the cosine of the angle 
  between the \B\ thrust and the $z$ axis.

  \item The scalar sum of the transverse momenta \pt\ in the ROE.  This
   sum includes neutral and charged particles.
\end{itemize}
The distributions of these input variables are shown in Fig.~\ref{fig:nnoinputvars}. 
Figure~\ref{fig:nnotraining} shows \nno\ for signal MC simulated events and continuum background samples 
(off-peak data) and the efficiency for signal and \qq\ background as a function
of cut on the minimum value of \nno. We require \nno\ to be greater than $-0.4$.
Note that later, we use this variable in the maximum likelihood fit described in Section~\ref{sec:maximum}. 

%-------------------------------
% Figure : NNO input variables
%-------------------------------
%-------------------------------

The samples used to train the neural net were correctly reconstructed MC simulated events 
and off-peak data. To avoid over-training, we used an independent 
sample of these data ({\em{i.e.}} distinct from the sample used for the 
training) to check the performance of the network.  
The training is stopped when the separation between the signal and continuum background
is stable.

%-------------------------------
% Figure : More NN plots .... 
%-------------------------------
%-----------------------------------------

%\input{misrec}
\section{\boldmath Mis-reconstructed signal and selection efficiency}
\label{sec:efficiency}

Mis-reconstructed signal candidates, referred to as self-cross-feed (\SCF) signal, may 
pass the selection requirements even if one or more of the pions assigned 
to the $\rho^+\rho^-$ state belongs to the other \B\ in the event. These SCF 
candidates constitute 50.7\% (27.9\%) of the accepted longitudinally (transversely) 
polarized signal. 
The majority of \SCF\ events have both charged pions from the 
$\rho^+\rho^-$ final state, and unbiased \CP\ information. 
These correct (right) track \SCF\ events are denoted by RT \SCF.  There 
is a \SCF\ component (13.8\% of the signal) where at least one 
track in $B_{\rm rec}$ is from the ROE.  These wrong track (WT) events are used
to determine the signal yield \nsig\ and \ptrue\ but have biased \CP information, 
and are treated separately for the \CP result. 
A systematic error is assigned to the \CP results from this type of signal event.  
The total selection efficiency for longitudinally (transversely) polarized 
signal is 7.7\% (10.5\%).  

%\input{tagging}
%-----------------------
% Flavor tagging
%-----------------------
\section{\boldmath \B\ meson Flavor Tagging\label{sec:tagging}}

A key ingredient in the measurement of time-dependent \CP\ asymmetries is to determine whether
at the time of decay of the \Btag, corresponding to $\deltat=0$, the \Brec\ was a \Bz\ or a 
\Bzb. This `flavor tagging'
is achieved with the analysis of the decay products of the recoiling \B\ meson \Btag.
The overwhelming majority of \B\ mesons decay to a final state that is flavor-specific, i.e. 
only accessible from either a \Bz\ or a \Bzb, but not from both. The purpose of the flavor tagging 
algorithm is to determine the flavor of \Btag\ with the highest possible efficiency $\epsilon_{\rm tag}$ 
and lowest possible probability \mistag\ of assigning a wrong flavor to \Btag.  It is not necessary
to fully reconstruct \Btag\ in order to determine its flavor. 
In the presence of a finite mistag probability \mistag, the \CP\ asymmetry is reduced by a dilution
factor $1-2\mistag$, so Eq. (\ref{eq:deltatdistribution}) becomes
\begin{eqnarray}
f_{\pm}(\deltat) &=& \frac{e^{ -\left|\deltat\right| / \tau_{\Bz}}}{4\tau_{\Bz}} \bigg{\{} 1 \pm (1-2\mistag) \times \nonumber \\
 && \eta_f [S \sin(\deltamd\deltat) - \C\cos(\deltamd\deltat)]\bigg{\}}. \nonumber \\ \label{eq:fdeltat}
\end{eqnarray}
The figure of merit for the performance of the tagging algorithm is the effective tagging efficiency 
\begin{equation}
Q = \epsilon_{\rm tag} (1-2\mistag)^2,
\end{equation}
which is related to the statistical uncertainty $\sigma$ in the coefficients $S$ and $C$ through
\begin{equation}
\sigma \propto \frac{1}{\sqrt{Q}}.
\end{equation}
We use a neural network based technique~\cite{babarsin2betaprd,babar_sin2beta} that isolates
primary leptons, kaons and pions from \B\ decays to final states containing $D^*$ 
mesons, and high momentum charged particles from \B\ decays, to determine the flavor of the \Btag.
The output of this algorithm is divided into seven mutually-exclusive categories. 
These are (in order of decreasing signal purity) \lepton, \kaonone, \kaontwo, \kaonpion, 
\pion, \other\ and \notag.  The performance of this algorithm is determined using fully reconstructed 
neutral \B\ decays to $D^{(*)-}(\pi^+,\rho^+, a_1^+)$ and is summarized in Table~\ref{tbl:taggingeff}. 
The categories assigned correspond to events with leptons, kaons and pions in the decay products of \Btag.
The \notag\ category of events contain no flavor information and therefore carry no weight in the 
time-dependent analysis.  In addition, these events are not considered useful for the branching fraction
 measurement since they are dominated by continuum background.
The total $Q$ of this algorithm is $32.7 \pm 0.7\%$.
%---------------------------------
% Table : Tagging
%---------------------------------
\begin{table}[!ht]
\caption{Tagging efficiency $\epsilon$, average mistag fraction  $\mistag$, mistag 
fraction difference between \Bz\ and \Bzb\ tagged events $\Delta \mistag$, and $Q$ for \Bztorhoprhom\ events.}
\label{tbl:taggingeff}
\begin{center}
\begin{tabular}{l|cccc} \hline\hline\vspace{-0.3cm}\\
Category   & $\epsilon_{{\rm tag}}$ (\%) & $\omega$ (\%)  & $\Delta\omega$ (\%) &  $Q$ (\%) \\[3pt] \hline
\lepton    & 8.2 $\pm$ 0.1               & 3.2 $\pm$ 0.5  &$-$0.2 $\pm$ 0.8       & 7.2 $\pm$ 0.2 \\
\kaonone   & 11.3 $\pm$ 0.1              & 3.7 $\pm$ 0.7  & 1.1 $\pm$ 1.2       & 9.7 $\pm$ 0.3 \\
\kaontwo   & 17.3 $\pm$ 0.2              & 14.2 $\pm$ 0.7 &$-$0.9 $\pm$ 1.1       & 8.8 $\pm$ 0.3 \\
\kaonpion  & 13.4 $\pm$ 0.1              & 20.8 $\pm$ 0.8 & 0.5 $\pm$ 1.3       & 4.6 $\pm$ 0.3 \\
\pion      & 13.8 $\pm$ 0.2              & 30.6 $\pm$ 0.8 & 4.1 $\pm$ 1.3       & 2.1 $\pm$ 0.2 \\
\other     & 9.4 $\pm$ 0.1               & 40.1 $\pm$ 1.0 & 2.3 $\pm$ 1.5       & 0.4 $\pm$ 0.1 \\
\notag     & 26.8 $\pm$ 0.2              & 50.0 $\pm$ 0.0 & $-$                 & 0.0 $\pm$ 0.0 \\\hline
Total      &                             &                &                     &32.7 $\pm$ 0.7 \\\hline\hline
\end{tabular}
\end{center}
\end{table}
%---------------------------------

%\input{mlfit}
%-----------------------
% Maximum Likelihood
%-----------------------
\section{\boldmath Likelihood Fit Method\label{sec:maximum}}

On applying the selection criteria described above, we obtain a sample of 
\ndata\ events that enter the fit.
These events are dominated by backgrounds from \qqbar\ ($81.4\%$)
and \BB\ ($16.6$\%) events. The remaining 2\% of events are 
considered as signal.
We distinguish between the following components in the fit:
\begin{list}{(\roman{Lcount})}
%    inform the list command to use this counter
{\usecounter{Lcount}
%    set rightmargin equal to leftmargin
\setlength{\rightmargin}{\leftmargin}}
\item correctly reconstructed signal,
\item \SCF\ signal, split into RT and WT parts,
\item charm $\Bpm$ backgrounds ($b\to c$),
\item charm $\Bz$ backgrounds ($b\to c$),
\item charmless $\Bz$ backgrounds,
\item charmless $\Bpm$ backgrounds,
\item continuum background.
\end{list}
The dominant \B\ backgrounds come from components (iii) and (iv). The signal, 
continuum and \B\ background models are described in Sections~\ref{sec:signalmodel},
~\ref{sec:continuummodel}, and~\ref{sec:bbgmodel}, respectively.

We use an unbinned, extended maximum likelihood (ML) fit to extract 
\nsig, \ptrue, \slong, and \clong. 
The likelihood function incorporates the following previously 
defined eight discriminating variables to distinguish signal from the backgrounds: 
\mes, \DeltaE, \deltat, \nno, and the $m_k$ and $\cos\theta_k$ values of the two $\rho$ mesons, where $k=1,2$.
For each of the aforementioned components $j$ we construct a probability 
density function (PDF) that is the product of one-dimensional PDFs for 
each of the variables.  The PDFs do not account for all possible correlations
among the discriminating variables and we account for possible biases as a result
of neglecting these correlations as discussed in Section~\ref{sec:fitvalidation}.
For each event $i$, the PDFs can be written as
\begin{eqnarray}
\Pdf_j^i &=& \Pdf_j(\mes^i)\Pdf_j(\DeltaE^i)\Pdf_j(\deltat^i)\Pdf_j(\nno^i) \times  \nonumber \\
         & & \Pdf_j(\mvone^i)\Pdf_j(\mvtwo^i)\Pdf_j(\cos\theta_{1}^{i})\Pdf_j(\cos\theta_{2}^{i}).\,\,
\end{eqnarray}
We determine the PDFs for signal and each of the \B\ background components from MC distributions 
for each discriminating variable. The PDFs for the continuum background are determined from
\onpeak\ and \offpeak\ data.  The likelihood function is
\begin{equation}
\like = \frac{\exp(-\sum_j n_j)}{N!} \prod_{i}^{N} \sum_{j}n_j\Pdf_j^i,
\end{equation}
where $n_j$ are the event yields of hypotheses $j$ (determined from the ML fit) and $N$
is the observed number of events in the sample.  The normalized exponent takes into 
account Poisson fluctuations in the total number of events. 
We simultaneously fit events in the mutually-exclusive flavor tag categories listed in 
Table~\ref{tbl:taggingeff}, excluding events in the \notag\ category.  There are six continuum background yields, one for each flavor tag category, 
and we use a single yield for the signal and each of the \B\ backgrounds,
while accounting for the relative fractions of events expected in each of the flavor tag 
categories. The fit results are obtained by minimizing the value of $-\ln\like$.

\subsection{Signal model}
\label{sec:signalmodel}

The signal has contributions from true and \SCF\ longitudinally ($\Long$) and transversely (${\Tran}$) polarized events.
In addition to this, the longitudinally polarized \SCF\ signal PDF is further sub-divided into the two categories 
of mis-reconstructed signal: RT and WT \SCF\ signal, and all signal PDFs are sub-divided according to the \Btag\
flavor tag category.  The total signal PDF is given by
\begin{eqnarray}
\Pdf_{\rm signal} &=& \ptrue \bigg{(}[1-f_{\rm RT\, \SCF}^{\Long} - f_{\rm WT\, \SCF}^{\Long}]\Pdf_{\rm true}^{\Long} \nonumber \\
                  && + f_{\rm RT\, \SCF}^{\Long}\Pdf_{\rm RT\, \SCF}^{\Long} + f_{\rm WT\, \SCF}^{\Long}\Pdf_{\rm WT\, \SCF}^{\Long} \bigg{)} \nonumber \\
                  && + (1-\ptrue)([ 1 - f_{\rm \SCF}^{\Tran}] \Pdf_{\rm true}^{\Tran} + f_{\rm \SCF}^{\Tran}\Pdf_{\rm \SCF}^{\Tran}),  \nonumber \\
\end{eqnarray}
where $f_{\rm RT(WT)\, \SCF}^{\Long}$ is the fraction of longitudinally polarized RT(WT) \SCF\ signal
and $f_{\rm \SCF}^{\Tran}$ is the fraction of transversely polarized \SCF\ signal.  The 
PDFs $\Pdf_{\rm RT(WT)\, \SCF}^{\Long}$ and  $\Pdf_{\rm \SCF}^{\Tran}$ are defined accordingly.
In order to extract
the observable \ptrue\ from the fit, we account for the different reconstruction efficiencies for 
longitudinally and transversely polarized signal using
\begin{eqnarray}
 \ptrue = \frac{\ptrue^{\rm obs}}{R + \ptrue^{\rm obs}(1-R)},
\end{eqnarray}
where $\ptrue^{\rm obs}$ is the fraction of signal from longitudinally 
polarized events in the data, and $R$ is the ratio of 
longitudinal to transverse polarized signal efficiencies.

All of the \SCF\ PDFs are further sub-divided into parts that contain correctly reconstructed and 
mis-reconstructed $\rho$ mesons.  Four sub-divisions are used, corresponding to both mesons being
correctly reconstructed (\TT), one being correctly reconstructed and the other being mis-reconstructed
(\TF\ or \FT), and both mesons being mis-reconstructed (\FF).  Here the first (second) $\rho$ meson is
positively (negatively) charged. The \SCF\ PDFs have the form
\begin{eqnarray}
\Pdf &=& \Pdf(\nno)\Pdf(\deltat) \large[ \nonumber \\
  &    & f_{\TT}\Pdf_{\TT}(\mes, \de, \coshelone, \mvone, \cosheltwo, \mvtwo) + \nonumber \\
  &    & f_{\TF}\Pdf_{\TF}(\mes, \de, \coshelone, \mvone, \cosheltwo, \mvtwo) + \nonumber \\
  &    & f_{\FT}\Pdf_{\FT}(\mes, \de, \coshelone, \mvone, \cosheltwo, \mvtwo) + \nonumber \\
  &    & f_{\FF}\Pdf_{\FF}(\mes, \de, \coshelone, \mvone, \cosheltwo, \mvtwo) \large], \nonumber \\
\label{eq:signalmesdemvcoshelpdf}
\end{eqnarray}
where $T$ and $F$ are correctly and mis-reconstructed $\rho$ mesons, respectively, 
 $f_{\TT}$, $f_{\TF}$, $f_{\FT}$, $f_{\FF}$ are the fractions of each type of mis-reconstructed event,
and the PDFs for \mes, \de, and the helicity and mass for each reconstructed $\rho$ meson are the products of 
one dimensional PDFs. The signal PDFs for \nno\ and \deltat\ have different shapes according to the \Btag\ 
flavor tag category assigned to an event. 

The longitudinally polarized true and RT \SCF\ signal \mes\ distribution are described 
by a Gaussian with a power law tail~\cite{cba,cbb,cbc} which takes the following form 
\begin{eqnarray}
f(m) & \propto& \exp\left(-\frac{(m-m_0)^2}{2\sigma^2}\right),\; m > m_0 - \xi\sigma, \nonumber \\
                            &\propto& \dfrac{n^n\,\exp(-\xi^2/2)} {\left( \frac{\xi(m_0 - m)}{\sigma} + n - \xi^2\right)^n},\; m \leq m_0 - \xi\sigma,
\end{eqnarray}
where $m_0$ is the mean, $\sigma$ is the width of the Gaussian part, and the parameters $\xi$ and $n$ describe the exponential tail.
The remaining PDFs for the signal \mes\ distributions are the sum of a Gaussian and a Gaussian with an exponential tail.
The signal \de\ distributions for the longitudinally polarized \SCF\ events are described by first and second
order polynomials, with the exception of \TT\ events, which are modeled with the sum of a second 
order polynomial and a Gaussian.  The other \de\ PDFs are described by the sum of Gaussian and a Gaussian with 
an exponential tail.  The \nno\ distributions are modeled using smoothed histograms of MC simulated events.  The signal
\deltat\ distribution is described by Eq. (\ref{eq:fdeltat}) convolved with a triple Gaussian resolution function
given by
\begin{eqnarray}
   {\cal R}_{\rm sig}(\dt,\sigma_{\dt}) &=& f_{\rm core}G\left(\dt,\mu_{\rm core}\sigma_{\dt}, \sigma_{\rm core} \sigma_{\dt}\right) + \nonumber \\
    &&  f_{\rm tail}G\left(\dt,\mu_{\rm tail}\sigma_{\dt}, \sigma_{\rm tail} \sigma_{\dt}\right) + \nonumber \\
    && f_{\rm outlier}G\left(\dt,\mu_{\rm outlier}, \sigma_{\rm outlier}\right)~,\label{eqn:signalresolutionfunction}
\end{eqnarray}
where G is a Gaussian with means $\mu_l$ and width $\sigma_l$ for $l=\rm core, tail$, and $\rm outlier$.
The parameters $\sigma_{\rm tail}$, $\sigma_{\rm outlier}$ and $\mu_{\rm outlier}$ are set to 3.0 \ps, 8.0 \ps\ and 0.0 \ps, respectively.
The remaining parameters of the resolution function are determined from signal MC simulated events scaled by
the differences between data and MC simulated events for large samples of exclusively reconstructed \B\ decays
to $D^{(*)-}(\pi^+,\rho^+, a_1^+)$ final states described in Section~\ref{sec:tagging}.  The values of the 
mean and width of the core Gaussian contribution to the resolution function depend on the flavor tagging category of an event.
There are three signal \deltat\ PDFs, one for the true and RT \SCF\ longitudinally polarized signal, one for the 
WT \SCF\ longitudinally polarized signal and one for the transversely polarized signal.  Each of these PDFs
has distinct values of $S$ and $C$ as described in Sections~\ref{sec:introduction} and~\ref{sec:efficiency}.  
The WT \SCF\ longitudinally polarized signal has a different resolution function with respect to the rest of the signal.
The \coshel\ distribution
for true signal events is described by the expected cosine squared or sine squared distribution multiplied by a polynomial acceptance
function. The \coshel\ PDFs for $T$ $\rho$ mesons in longitudinally (transversely) polarized \SCF\ signal are described by sixth order 
polynomials (smoothed histograms of MC simulated events).  The \coshel\ PDFs for mis-reconstructed $\rho$ mesons
are described by smoothed histograms of MC simulated events. The longitudinally polarized signal $\rho$ mass PDFs 
are described using relativistic Breit-Wigners and third order polynomials for $T$ and $F$ $\rho$ contributions, 
respectively.  The transversely polarized signal \SCF\ $\rho$ mass PDF distributions are described using smoothed histograms 
of MC simulated events.

In addition to \nsig, \ptrue, \slong, and \clong, we determine the mean of the core 
Gaussian part of the \mes\ and \de\ distributions from the fit.  We obtain parameters 
consistent with expectations from MC simulated events.

\subsection{Continuum background model}
\label{sec:continuummodel}

The continuum background PDFs are also sub-divided according to the mis-reconstruction of $\rho$ mesons 
in the final state and have the form
\begin{eqnarray}
\Pdf &=& \Pdf(\mes)\Pdf(\de)\Pdf(\nno)\Pdf(\deltat) \bigg{[} \nonumber \\
  &    & f_{\TT}\Pdf_{\TT}(\coshelone, \mvone, \cosheltwo, \mvtwo) + \nonumber \\
  &    & f_{\TF}\Pdf_{\TF}(\coshelone, \mvone, \cosheltwo, \mvtwo) + \nonumber \\
  &    & f_{\FT}\Pdf_{\FT}(\coshelone, \mvone, \cosheltwo, \mvtwo) + \nonumber \\
  &    & f_{\FF}\Pdf_{\FF}(\coshelone, \mvone, \cosheltwo, \mvtwo) \bigg{]}, \label{eq:contmesdemvcoshelpdf}
\end{eqnarray}
where helicity and mass PDFs for each reconstructed $\rho$ meson are the products of 
one dimensional PDFs.

The continuum distribution for \mes\ is described by a phase-space-motivated
distribution~\cite{argus2} with the following form
\begin{equation}
f(x) \propto x\sqrt{1 - x^2}\,\cdot\,\exp[ \xi\,(1 - x^2) ]\,\cdot\, \theta(\mes), \label{eq:argusfunction}
\end{equation}
where $\theta(\mes) = 1$ for $\mes \leq \sqrt{s}/2$ and $\theta(\mes) = 0$ for $\mes > \sqrt{s}/2$ and
$x=2\mes/\sqrt{s}$.
The \de\ and \nno\ shapes are
modeled with  third and fourth order polynomials, respectively.  The
parameters of the \mes, \de\ and \nno\ shapes are allowed to vary in the fit 
to the \onpeak\ data. The continuum $\rho$ mass distribution is described using a
relativistic Breit-Wigner and a third order polynomial PDF for $T$ and $F$ 
$\rho$ contributions, respectively.  The \coshel\ distribution is 
described by a  third order polynomial.  The continuum
\deltat\ distribution has a prompt lifetime component convolved with a triple 
Gaussian resolution function. The parameters of the $\rho$ mass and helicity
distributions are obtained from a fit to the \offpeak\ data, and the remaining parameters
are determined in the fit. 

\subsection{\B\ background model}
\label{sec:bbgmodel}

\subsubsection{Charm \B\ backgrounds}

Combinatorial events from $b \to c$ transitions are the dominant \B\ backgrounds.  These 
components have shapes similar to continuum and do not peak in the signal region for the
discriminating variables.  The functional form used for the PDFs of these background components
is given by Eq. (\ref{eq:contmesdemvcoshelpdf}). We parameterize the \mes\ and \de\ distributions 
of these backgrounds using the phase-space-motivated distribution of Eq. (\ref{eq:argusfunction}), and a third 
order polynomial, respectively.  The $\rho$ mass distribution is described using a
relativistic Breit-Wigner and a third order polynomial PDF for $T$ and $F$ 
$\rho$ contributions, respectively. The remaining PDFs are described using smoothed histograms
of MC simulated events. Each of the PDF parameters for these backgrounds are determined from samples of MC 
simulated events, and the yields of these components are determined in the fit.
When studying systematic uncertainties we modify the PDF used for \deltat\ so that it
has a form similar to the signal which uses the resolution function of Eq. (\ref{eqn:signalresolutionfunction}).
An effective lifetime that is smaller than $\tau_{\Bz}$ is used to account for 
mis-reconstruction of these events and the finite charm meson lifetime.  The value of this parameter
is obtained by fitting MC simulated events for this category of events.

\subsubsection{Charmless \B\ backgrounds}

Some of the charmless \B\ backgrounds have PDFs similar to the signal for 
one or more of the discriminating variables, so it is important to correctly account 
for such events in the fit.  We consider the 20 exclusive and 2 inclusive components of
this type of \B\ background listed in Table~\ref{tbl:bbackground}.  
If a charmless \B\ background decay contributes an event yield more than 1\% of the expected
signal yield, we model that mode exclusively.  The remaining 140 charmless background components considered 
were combined with the appropriate weightings to form the neutral and charged inclusive charmless \B\ background 
components.
Where possible we use branching fractions from existing measurements.  Where measurements are not available,
we have either tried to use \su{2} and \su{3} flavor symmetries to relate an unmeasured decay to a measured one, or 
where this is not possible, we have assigned a branching fraction of 10\e{-6} to a decay mode. 
An uncertainty of 100\% is assumed on all extrapolated branching fractions.  We assume that the
$a_1$ meson decays into a three pion final state via $\rho \pi$.  The decay $\Bz \to a_1^0 \pi^0$ is
penguin dominated and is expected to have a small branching fraction, as is the case for the 
penguin dominated decays $\Bz \to \rho^0 \pi^0$ and $\Bz \to \rho^0 \rho^0$.  As a result, we have neglected possible
contributions from this potential source of background. 
Although we don't see evidence for $\Bz\to\rho^\mp\pi^\pm\piz$, we do consider this channel as a potential source of background. 
Contribution from $\Bz\to{4}\pi$ is expected to be even smaller and is neglected.
When considering the systematic uncertainty arising from interference between signal and other
$\pi^{+}\pi^{-}\pi^{0}\pi^{0}$ final states, we assume that the non-resonant $\pi^{+}\pi^{-}\pi^{0}\pi^{0}$ final state
has the same branching fraction as that of $\Bz\to\rho^\mp\pi^\pm\piz$, where $\br(\Bz\to\rho^\mp\pi^\pm\piz)$
is calculated from the yield determined in the $\Bz \to \rho^+ \rho^-$ nominal fit. This is discussed
further in the Appendix.      

\begin{table}
\caption{The components of charmless \B\ backgrounds considered, along with the 
  branching fraction (${\cal B}$) and number of events ($N_{bg}$) expected in the selected data sample,
  where $\dag$ indicates a longitudinally polarized final state, and \NonRes\ denotes
  a non-resonant final state.  Where appropriate, branching fractions of these decays 
  are quoted, including the branching fraction of sub-decay modes and measured or expected values of \ptrue.}\label{tbl:bbackground}
\begin{tabular}{l|rr}\hline\hline\noalign{\vskip1pt}
Decay Mode                                         & ${\cal B} \e{-6}$   & $N_{bg}$ \\[3pt] \hline\noalign{\vskip1pt}
\classzero\cite{ref:bchh,ref:bchi}                 & 55 $\pm$ 3          & 11 $\pm$ 1 \\
\classone                                          & 10 $\pm$ 10         & 15 $\pm$ 15 \\
\classtwo\cite{ref:bchc,ref:bchd,ref:bche}         & 11 $\pm$ 2          & 73 $\pm$ 10 \\
\classthree\cite{babarrhoprhoz,bellerhorho0} $\dag$& 17 $\pm$ 3          & 71 $\pm$ 12\\
\classfour        & 10 $\pm$ 10                                      &  9 $\pm$ 8 \\
\classfive        & 10 $\pm$ 10                                      &  9 $\pm$ 9 \\
\classsix\cite{ref:bchh,ref:bchi,ref:bche}         & 11 $\pm$  1     &  9 $\pm$ 1 \\
\classseven       & 20 $\pm$ 20                                      & 60 $\pm$ 60 \\
\classeight       & 10 $\pm$ 10                                      & 12 $\pm$ 12 \\
\classnine $\dag$ & 8  $\pm$  8                                      & 13 $\pm$ 13 \\
\classten         & 20 $\pm$ 20                                      & 49 $\pm$ 49 \\
\classtwelve      &    $-$                                           & 59 $\pm$ 15 \\
\classthirteen    & 30 $\pm$ 3                                       & 23 $\pm$ 23 \\
\classfourteen\cite{ref:bchk,ref:bche}    & 24 $\pm$ 3               & 42 $\pm$  4 \\
\classfifteen     & 23 $\pm$ 23                                      & 15 $\pm$ 15 \\
\classsixteen\cite{ref:bchn}     & 20 $\pm$  2                       &  7 $\pm$  1 \\
\classseventeen   & 10 $\pm$ 10                                      & 45 $\pm$ 45\\
\classeighteen    & 10 $\pm$ 10                                      &  8 $\pm$  8\\
\classnineteen\cite{ref:bchp} $\dag$ & 16 $\pm$ 16                   & 43 $\pm$ 43\\
\classtwenty\cite{ref:bchp}   $\dag$ & 16 $\pm$ 15                   &  9 $\pm$ 8\\
\classtwentyone \cite{ref:bchn}   & 40 $\pm$  4                      & 102 $\pm$ 9\\
\classtwentythree &                 $-$                              & 88 $\pm$ 22 \\[3pt] \hline\hline
\end{tabular}
\end{table}

The functional form used for the PDFs of these charmless \B\ background components is given by Eq. (\ref{eq:contmesdemvcoshelpdf}).  
The \de\ distributions are described by third order polynomials, except 
for non-resonant $\Bz\to\rho^\mp\pi^\pm\piz$ which uses smoothed histograms of MC simulated events.  The 
$m_\rho$ distributions for true $\rho$ mesons are parameterized using a relativistic 
Breit-Wigner, and the fake $m_\rho$ (combinatorial $\pi^\pm\piz$) distribution is 
described using  third order polynomials.  The remaining background shapes are 
described using smoothed histograms of MC simulated events.
The yield for \classseventeen\ decays is allowed to vary in the fit.  All other charmless 
background yields are fixed to expectations.  This constraint is relaxed when studying 
possible sources of systematic uncertainties.
When studying systematic uncertainties from possible \CP\ violation in the \B\ background, we 
modify the PDF used for \deltat\ so that it has a form similar to the signal 
which uses the signal resolution function in Eq. (\ref{eqn:signalresolutionfunction}).  

%\input{fitvalidation}
%----------------------------------
% Maximum Likelihood Fit Validation
%----------------------------------
\section{\boldmath Likelihood Fit Validation\label{sec:fitvalidation}}

Before applying the fitting procedure to the data, we subject it to various tests.
The aim of these tests are to verify that one can extract the signal observables 
\nsig, \ptrue, \slong, and \clong\ in a controlled way.  Consistency of the 
likelihood fit is checked with ensembles of experiments simulated from the PDFs.
The event yields are generated according to a Poisson distribution with mean 
$n_j$ for each category in these ensembles.
In each of these tests we verify that the values generated for the signal 
observables are reproduced with the expected resolution. 
The distribution of $-\ln\like$
for the ensemble of experiments in comparison to that obtained when fitting 
the data provides an additional, but limited, cross-check of the consistency.

The PDFs used in the likelihood do not account for all possible correlations
among the discriminating variables.  We account for possible biases as a result
of neglecting these correlations by fitting ensembles of experiments obtained 
from samples of signal and the charmless \B\ background MC simulated events 
combined with charm backgrounds and \qqbar\ background events simulated from the PDFs. 
The MC simulated events used in these ensembles have these correlations modeled. 
We find a positive bias of \fitbiasonyield\ events on \nsig, and a negative bias of
\fitbiasonfl\ on \ptrue, and we do not observe a significant bias on \slong\ and
\clong.  

As continuum events are the dominant background, we apply the fitting procedure 
to the \offpeak\ data (after correcting for the difference in $\sqrt{s}$) to confirm
that we do not find a fake signal in this control sample of events. We fit 790 \offpeak\
events and extract signal and continuum yields of $8 \pm 7$ and $782 \pm 28$ events, respectively.

A blind analysis technique has been used for the extraction of \slong\ and \clong, where 
the actual values of these observables have been hidden by an offset.  With the values
of \slong\ and \clong\ hidden we perform the following fit cross-checks.
We first verify that the uncertainties on the signal observables, and the value of $-\ln\like$ 
obtained from the fit to data, are compatible with the ensembles of simulated experiments described above.
We then validate the stability of our results by introducing a variety of modifications 
to the fitting procedure.  In addition to the nominal set of variables determined from data, we 
allow the physics parameters $\tau_{\Bz}$ and $\dm$ to vary in turn.  The signal observables do not
change significantly when doing this, and the results obtained for $\tau_{\Bz}$ and $\dm$
are $1.72 \pm 0.16$ \ps\ and $0.36 \pm 0.22 \, \hbar/\ps$, respectively,
consistent with the reported world average~\cite{pdg2006}.  

The \lepton\
and \kaonone\ tagged events have the highest signal purity, and lowest \mistag\ probability, and dominate our \CP\ results.  
We check that the results obtained from fitting only these categories of events are compatible
with our nominal blind results. When doing this, we observe shifts of $-0.08$ ($-0.05$) 
on \slong\ (\clong) relative to the nominal result. 

Given that there are a number of
\B\ backgrounds that contribute to the data that have yields fixed to expectations, we 
validate this assumption by allowing each fixed \B\ background yield to vary in the fit.
We obtain background yields consistent with our expectations, and observe the shifts on signal 
parameters listed in Table~\ref{tab:fit:floatbackgrounds}. A systematic
uncertainty corresponding to the largest observed deviation is assigned on 
our results.

Once these checks have been completed, the fitting procedure is frozen and we extract
the values of \slong\ and \clong\ by removing the offset.

\begin{table}[!ht]
  \begin{center}
  \caption{The shifts on signal yield ($\delta N(\mathrm{signal})$), 
           fraction of longitudinally polarized events ($\delta\ptrue$), 
           \slong\ ($\delta\slong$), and \clong\ ($\delta\clong$) 
           obtained when floating the yield of each fixed \B\ background 
           in turn. The symbol
           $\dag$ indicates a longitudinally polarized final state, 
           and \NonRes\ denotes a non-resonant final state.  The 
           yields of decay modes not listed in this table 
           are allowed to vary in the nominal fit.}
  \label{tab:fit:floatbackgrounds}
    \begin{tabular}{l|rrrr}\hline\hline
      Decay Mode & $\delta N(\mathrm{signal})$ & $\delta\ptrue$ & $\delta\slong$ & $\delta\clong$ \\ [3pt]\hline\noalign{\vskip1pt}
      \classzero         & 3     & $-$0.002    &  $-$0.002 &  $-$0.003     \\
      \classone          & 11    & 0.001     &  0.002  &  $-$0.001    \\
      \classtwo          & 9     & 0.000     &  0.001  &  $-$0.002     \\
      \classthree $\dag$ & $-$8  & $-$0.001    &  $-$0.002 &  0.000    \\
      \classfour         & 1     & 0.000     &  $-$0.001 &  $-$0.001     \\
      \classfive         & $-$5  & 0.001     &  $-$0.001 &  $-$0.000    \\
      \classsix          & $-$13 & $-$0.001    &  $-$0.002 &  0.001     \\
      \classseven        & 17    & 0.003     &  0.006  &  $-$0.001     \\
      \classeight        & $-$2  & 0.000     &  0.001  &  $-$0.002     \\
      \classnine $\dag$  & $-$34 & $-$0.006    &  $-$0.028 &  $-$0.003     \\
      \classten          & 5     & 0.001     &  0.004  &  $-$0.001     \\
      \classtwelve       & $-$3  & $-$0.001    &  $-$0.002 &  0.000    \\
      \classthirteen     & $-$4  & 0.000     &  0.000  &  0.000    \\
      \classfourteen     & $-$5  & 0.000     &  $-$0.001 &  0.000    \\
      \classfifteen      & $-$25 & $-$0.003    &  $-$0.009 &  $-$0.001     \\
      \classsixteen      &  1    & $-$0.002    &  $-$0.001 &  $-$0.002     \\
      \classeighteen     & $-$31 & $-$0.003    &  $-$0.009 &  $-$0.003    \\
      \classnineteen $\dag$  & $-$25 & $-$0.002    &  $-$0.025 &  $-$0.004    \\
      \classtwenty   $\dag$  &  11 & 0.000     &  0.006  &   0.000    \\
      \classtwentyone    & 10    & 0.002     &  0.003  &   $-$0.002    \\
      \classtwentythree  & 1     & 0.000     &  0.000  &  $-$0.001    \\ [3pt]\hline\hline
    \end{tabular}
  \end{center}
\end{table}

%\input{results}
%-----------------------
% Results
%-----------------------
\section{\boldmath Results\label{sec:results}}

\subsection{Fit results}

\begin{figure}[!ht]
\begin{center}
\resizebox{9cm}{!}{
 \includegraphics{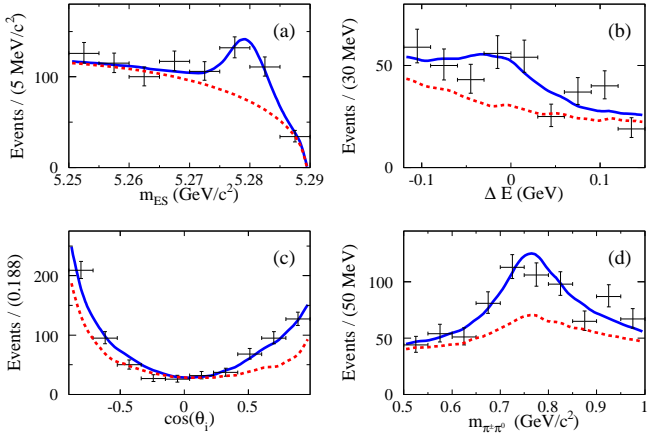}
}
\caption{The distributions for the highest purity tagged events for
the variables (a) \mes,  (b) \DeltaE, (c) cosine of the $\rho$ helicity
angle and  (d) \mv.  The dashed lines are the sum of backgrounds, and the 
solid lines are the full PDF.
} \label{fig:plots}
\end{center}
\end{figure}

\begin{figure}[!ht]
\begin{center}
\resizebox{9cm}{!}{
 \includegraphics{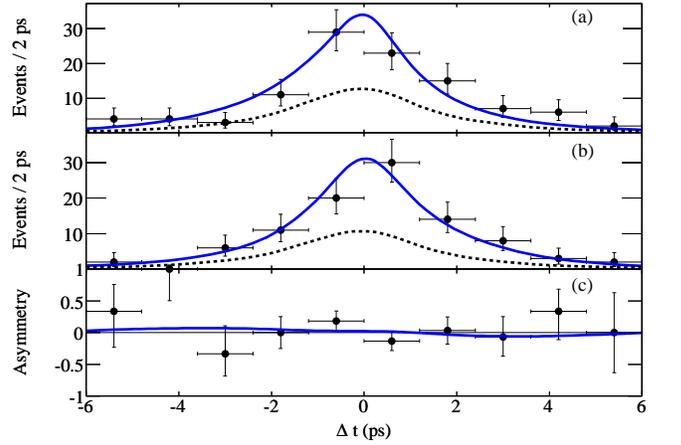}
}
 \caption{The \deltat\ distribution for a sample of events enriched in signal for
  (a) $\Bz$ and (b) $\Bzb$ tagged events. The dashed lines are the sum of backgrounds,
  and the solid lines are the sum of signal and backgrounds.  The time-dependent \CP asymmetry (see text) is shown in (c), where
  the curve is the measured asymmetry.} \label{fig:dtplots}
\end{center}
\end{figure}

From the ML fit described above,
we obtain the following results
\begin{eqnarray}
N({\rm signal}) &=& \correctedsignalyield,\nonumber \\
\ptrue &=& \correctedfl,\nonumber \\
\slong &=& \correctedslong,\nonumber \\
\clong &=& \correctedclong,\nonumber
\end{eqnarray}
after correction for a $+\fitbiasonyield$ event fit bias (see Section~\ref{sec:fitvalidation}), 
a $-$76 event bias from \SCF\ on the signal yield (see Section~\ref{sec:Systematics})
and a correction for a $-\fitbiasonfl$ fit bias on \ptrue.  
The correlation between \slong\ and \clong\ is \sccorrelation. 
We discuss the origin of these fit biases in Section~\ref{sec:Systematics}.
We calculate the branching 
fraction of this decay using $\br=\nsig/(\epsilon N_{\bbbar})$, where $\epsilon$
is the efficiency for signal corresponding to the observed \ptrue, and $N_{\B\ \rm pairs}$ is
the number of \BB\ pairs analyzed.  We obtain 
\begin{eqnarray}
\br(\Bztorhoprhom) &=& \correctedbf.\nonumber
\end{eqnarray}
The $\Bz\to\rho^\mp\pi^\pm\piz$ background yield obtained from the fit is $9.2 \pm 53.6$ events.
Figure~\ref{fig:plots} shows
distributions of \mes, \DeltaE, \coshel\ and \mv\ for the \lepton\ and \kaonone\
tagged events with a loose
requirement on \nno. Relative to the total number of events in the fit, 
the plot of \mes\ contains 15.6\% of the
signal and 1.1\% of the total background. For the other plots
there is an added constraint that $\mes > 5.27 \gevcc$; these
requirements retain 13.9\% of the signal and 0.4\% of the total background.
Figure~\ref{fig:dtplots} shows the \deltat\ distribution for \Bz and \Bzb
tagged events, as well as the time-dependent decay-rate asymmetry of Eq. (\ref{eq:timedepasym}).
Here we apply the same selection criteria as in Fig.~\ref{fig:plots}(b)-(d).

\subsection{Systematic uncertainty studies}
\label{sec:Systematics}

Table~\ref{tbl:systsummary} lists the possible sources of systematic uncertainties 
on the values of the \nsig, \ptrue, \slong, and \clong\ that have been studied,
and are described in the following.

\begin{table*}[!ht]
\caption{Summary of additive systematic uncertainty contributions on the signal yield \nsig, \ptrue, \slong\ and \clong.}\label{tbl:systsummary}
\begin{center}
\begin{tabular}{l|cccc}\hline\hline \vspace{-0.3cm}\\
Contribution                 & $\sigma(\nsig)$ & $\sigma(\ptrue)$ & $\sigma(\slong)$ & $\sigma(\clong)$  \\[3pt] \hline\noalign{\vskip1pt}
%%%%%%%%%%%%%%%%%%%%%%%%%%%%%%%%%%%%%%%%%%%%%%%%%%%%%%%%%%%%%%%%%%%%%%%%%%%%%%%%%%%%%%%%%%%%%%%%%%%%%%%%%%%%%%%%%%%%%%%%%% 
PDF parameterisation         & $\,^{+23}_{-41}$   & $\,^{+0.019}_{-0.005}$   & $\,^{+0.02}_{-0.04}$    & $0.03$ \\[6pt]
\SCF\ fraction               & 76                 & $0.003$                  & $0.00$                 & $0.02$ \\[6pt]
Control samples calibration  & 16                 & 0.006                    & 0.01                    & 0.01 \\[6pt]
$\mes$ endpoint              & 12                 & 0.001                    & 0.00                    & 0.01 \\[6pt]
\B\ background normalization & $\,^{+16}_{-20}$   & $\,^{+0.005}_{-0.002}$   & 0.01                    & $0.00$\\[6pt]
floating \B\ backgrounds     & 34                 & 0.006                    & 0.03                    & 0.00\\[6pt]
\B\ background \CP\ asymmetry& $ 2$            & $\,^{+0.001}_{-0.000}$   & $0.00$                     & $\,^{+0.02}_{-0.01}$ \\[6pt]
$\tau_B$                     & $\,^{+0}_{-1}$     & $\,^{+0.001}_{-0.000}$   & $0.00$                  & $0.00$ \\[6pt]
\dm                          & $\,^{+0}_{-1}$     & $\,^{+0.001}_{-0.000}$   & $0.00$                  & $0.00$\\[6pt]
tagging and dilution         & $\,^{+2}_{-16}$    & $\,^{+0.010}_{-0.001}$   & $0.00$                  & $0.01$\\[6pt]
transverse polarization \CP\ asymmetry & $\,^{+0}_{-9}$     & $\,^{+0.006}_{-0.000}$   & $0.01$                  & $0.01$\\[6pt]
Wrong track \SCF\ \CP\ asymmetry       & $\,^{+0}_{-3}$     & $\,^{+0.001}_{-0.000}$   & $0.01$                  & $0.01$\\[6pt]
DCSD decays                  & $-$                & $-$                      & 0.01                    & 0.04 \\[6pt]
Interference                 & 18                 & 0.000                    & 0.01                    & 0.01 \\[6pt]
Fit Bias                     & 19                 & 0.008                    & 0.02                    & 0.02 \\[6pt]
SVT Alignment                & $-$                & $-$                      & 0.01                    & 0.01 \\[6pt] \hline \\[-6pt]
%%%%%%%%%%%%%%%%%%%%%%%%%%%%%%%%%%%%%%%%%%%%%%%%%%%%%%%%%%%%%%%%%%%%%%%%%%%%%%%%%%%%%%%%%%%%%%%%%%%%%%%%%%%%%%%%%%%%%%%%%%
Total                        & $^{+94}_{-102}$     & $^{+0.03}_{-0.01}$       & $^{+0.05}_{-0.06}$      & $ 0.06$ \\ [2pt] 
\hline \hline
\end{tabular}
\end{center}
\end{table*}

\begin{itemize} 
\item The uncertainty from PDF parameterisation is obtained by varying PDF shape parameters by 
      $\pm 1 \sigma$, in turn.  The deviations obtained are added in quadrature
      to give the quoted uncertainty.
\item The systematic uncertainty from the fraction of \SCF\ events is obtained 
      from the difference between the nominal fit result and that obtained when fitting the data
      and removing the \SCF\ from the fit model.  In the case of the signal yield we correct
      for half of the difference observed, and use 100\% of the correction as an uncertainty.
      The uncertainty on the other signal observables comes from the difference observed between
      the two sets of fit results.
\item The kinematic endpoint position in \mes\ is extracted from the fit.  Changes in 
      beam energy in the data can affect the endpoint position.  To account for possible
      effects of this, we vary the kinematic endpoint position in \mes\ by $\pm 0.45 \mevcc$ 
      which is determined from control samples of fully reconstructed $\Bz$ decays.
\item The uncertainty from the \mes\ and \de\ widths is obtained from the
      observed shifts relative to our nominal result, when allowing these parameters to vary 
      independently in the fit to data.
\item We vary the \B\ background normalization within expectations for each background in turn.  The
      deviations obtained are added in quadrature
      to give the quoted uncertainty from this source.
\item As the branching fractions of some of the \B\ backgrounds are not well known, we assign an additional 
      uncertainty coming from the maximum shifts obtained when allowing each of the fixed backgrounds
      to vary in turn in the fit to data.  
\item Additional uncertainties on the \CP\ results come from possible \CP\ violation in the \B\ background.  
      We replace the \deltat\ PDFs used by each of the \B\ backgrounds in turn by one resembling the signal.
      Charged \B\ backgrounds can have non-zero values of \Acp, and neutral \B\ backgrounds can have non-zero
      values of $S$ and $C$. We use existing experimental constraints where possible, 
      otherwise we allow for a \CP asymmetry up to 10\% in \B\ decays 
      to final states with charm, and up to 50\% in \B\ decays to charmless final states.  
\item The physics parameters $\tau_{\Bz}=1.530\pm 0.009$ \ps\ and $\deltamd = 0.507 \pm 0.005$ $\hbar/\ps$~\cite{pdg2006} 
      are varied within the quoted uncertainty.  
\item The tagging and mistag fractions for signal and the \B\ backgrounds are corrected for data/MC 
      differences observed in samples of fully reconstructed hadronic \B\ decays.  Each of the tagging 
      and mistag parameters is varied in turn by the uncertainty from the correction.
      The deviations obtained are added in quadrature
      to give the quoted uncertainty from this source.
\item Allowing for possible \CP\ violation in the transverse polarization, and in the WT 
      longitudinally polarized signal \SCF\ events results in additional uncertainties on signal observables.
      We vary $S$ and $C$ by $\pm 0.5$ ($\pm 1.0$) for the transverse polarization (WT \SCF).
\item Possible \CP\ violation from interference in doubly Cabibbo-suppressed decays (DCSD) on the tag side of the 
      event~\cite{ref:dcsd} contribute to systematic uncertainties on \slong\ and \clong.  
\item We estimate the systematic error on our results coming from neglecting the interference between 
      \Bztorhoprhom\ and other $4\pi$ final states: $\B \to a_1\pi$, $\rho \pi\pi^0$ and $\pi\pi\piz\piz$. 
      Details of this calculation are given in the Appendix.
\item As the PDFs used in the ML fit do not account for all of the correlations between discriminating 
      variables used in the fit, the results have a small bias.  We calculate the fit bias on the signal 
      observables as described in Section~\ref{sec:fitvalidation}.
      The observed bias on \nsig, and \ptrue\ is corrected, and 100\% of the correction is assigned 
      as a systematic uncertainty.  We do not observe a significant bias on \slong\ and \clong, and 
      conservatively assign a systematic uncertainty from fit bias based on the statistical precision 
      of this test.
\item Small imperfections in the knowledge of the geometry of the SVT over time can affect the 
      measurement of \slong\ and \clong.  We vary the alignment according to the results obtained from the 
      study of $\epem \to e^+e^-$, $\mu^+\mu^-$ events in order to estimate the magnitude of this systematic error
      on our \CP\ results.
\end{itemize}

The branching fraction has multiplicative systematic uncertainties from 
the reconstruction of \piz\ mesons in the detector (6\%), uncertainties
in the reconstruction of charged particles (0.8\%), and the discrimination
of $\pi^\pm$ from other types of charged particles (1\%).  In addition to these 
uncertainties, there is a 1.1\% uncertainty on the number of \bb\ pairs in the
data sample.  The statistical uncertainties arising from the MC samples 
used in this analysis are negligible.

%\input{alpha}
%-----------------------
% Measurement of alpha
%-----------------------
\section{\boldmath Constraints on the Unitarity Triangle\label{sec:alpha}}
%\begin{itemize}
%\item{Isospin analysis - formalism and results}
%\item{\su{3} model formalism and results}
%\end{itemize}
\subsection{\boldmath The \su{2} isospin analysis}

\begin{figure}[!h]
\begin{center}
  \resizebox{6.5cm}{!}{\includegraphics{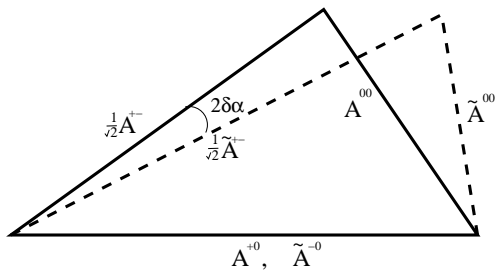}}
\end{center}
 \caption{The isospin triangle for $\B\to\rho\rho$ decays.}
\label{fig:ispintriangle}
\end{figure}

In \su{2}\ isospin symmetry, the amplitudes of \B\ decays to $\rho\rho$ final states~\cite{gronaulondon}
are related by: 
\begin{eqnarray}
\frac{1}{\sqrt{2}}A^{+-}=A^{+0}-A^{00},\label{eq:iaone}\\
\frac{1}{\sqrt{2}}\overline{A}^{+-}=\overline{A}^{-0}-\overline{A}^{00},\label{eq:iatwo}
\end{eqnarray}
for the longitudinal polarization and each of the \CP\ eigenstates of the transverse polarization, 
where $A^{ij}$ ($\overline{A}^{ij}$) are the amplitudes of $\B(\overline{B})$ 
decays to the final state with charge $ij$, where $i = +,-,0$ and $j = -,0$. These
two relations correspond to triangles in a complex plane as shown in 
Fig.~\ref{fig:ispintriangle}.  In the usual phase convention~\cite{gronaulondon}, the amplitudes $\bar A$ are 
rotated to $\tilde A$ in order to align the base of the triangles.
After aligning $A^{+0}$ and $\overline{A}^{-0}$, the
phase difference between $A^{+-}$ and $\overline{A}^{+-}$ is
$2\delta\alpha$. There are two sources of ambiguities on the measurement of $\alpha$. 
There is a four-fold ambiguity coming from the orientation of the isospin triangles,
and a two-fold ambiguity from the trigonometric conversion in Eq. (\ref{eq:seff}).   
In order to measure $\alpha$ one must measure the
branching fractions and charge asymmetries of
\B\ decays to $\rho^+\rho^-$, $\rho^\pm\rho^0$, $\rho^0\rho^0$.

There are several assumptions implicitly used in the isospin-based
direct measurements of $\alpha$:
\begin{itemize}
 \item This approach only considers tree and gluonic penguin
   contributions. Possible contributions from electroweak penguins
   (EWP) are neglected as they do not obey \su{2}\ isospin symmetry.
   These EWPs have the same topology as the gluonic penguin diagram in
   Fig.~\ref{fig:feynmangraphs}, with the gluon replaced by
   $\gamma$ or $Z^0$ bosons. In the absence of EWP contributions
   $|A^{+0}|=|\overline{A}^{-0}|$, and \Acp\ is zero for 
   $\B^+ \to \rho^+\rho^0$. 
   Several groups have estimated the correction due to the
   \su{2}\ breaking effect of EWP contributions to be $1.5-2.0$
   degrees~\cite{electroweak_pengin_calculation,zupanewp}. These
   estimates consider contributions from the two EWP operators
   assumed to be dominant in the effective Hamiltonian.

 \item The possible effect on the isospin analysis from $\rho^0 - \omega$ mixing~\cite{zupanewp} is neglected.

 \item Other \su{2}\ symmetry breaking effects are neglected.  
   Estimates of the magnitude of these effects are much less than 
   the current experimental precision~\cite{gardner1999,gardner2005}.
   Possible isospin $5/2$ amplitudes also break the \su{2}\ 
   triangle construction~\cite{ref:london2006}.

 \item The isospin analysis outlined above neglects possible I=1 
   amplitudes~\cite{falk}.  The presence of I=1 amplitudes in 
   \Bztorhoprhom\ can be tested by measuring \slong\ and \clong\ 
   for different ranges of the invariant $\pi^\pm\pi^0$ 
   mass.
\end{itemize}

We constrain the CKM angle $\alpha$ and penguin contribution $\delta\alpha$ from an isospin
analysis of $B \to \rho\rho$ decays.  The inputs to the isospin analysis are the 
amplitudes of
the \CP-even longitudinal polarization of the $\rho\rho$ final state, as
well as the measured values of \slong\ and \clong\
for \Bztorhoprhom.  We use the following numerical inputs in the isospin analysis:
\begin{itemize}
 \item The measurements of ${\cal B}(\Bztorhoprhom)$, \ptrue, \slong, and \clong\ presented here.
 \item The measurements of ${\cal B}$ and \ptrue\ for $\Bp\to\rhop\rhoz$ from Ref.~\cite{babarrhoprhoz}.
 \item The measurement of ${ \cal B}(\Bz\to\rhoz\rhoz)$ from Ref.~\cite{babarrhozrhoz}.
\end{itemize}

\begin{figure}[!ht]
 \begin{center}
 \resizebox{8cm}{!}{
 %\rotatebox{-90.0}{ 
  \includegraphics{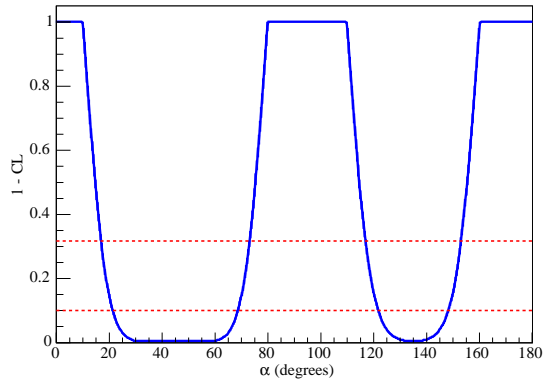} 
 %}
}
\caption{Confidence level on $\alpha$ obtained from the isospin analysis
with the statistical method described in~\cite{ckmfitter}.
The dashed lines correspond to the 68\% (top) and 90\% (bottom)
CL intervals. } \label{fig:alpha}
 \end{center}
\end{figure}

To interpret our results in terms of a constraint on $\alpha$ from
the isospin relations,  we construct a $\chi^2$ that includes
the measured quantities expressed as the lengths of the sides of
the isospin triangles and we determine the minimum $\chi^2_0$.
We have adopted a simulated-experiment technique to compute the confidence
level (CL) on $\alpha$; our method is similar to
the approach proposed in Ref.~\cite{FC98}.
For each value of $\alpha$, scanned between
$0$ and $180^\circ$, we determine the
difference $\Delta \chi^2_{{\rm DATA}}(\alpha)$  between  the minimum
of $\chi^2(\alpha)$ and $\chi^2_0$. We then generate MC experiments
around the central values obtained from the fit to data with
the given value of $\alpha$ and
we apply the same procedure. The fraction of these experiments in
which  $\Delta \chi^2_{{\rm MC}}(\alpha)$ is smaller than
$\Delta \chi^2_{{\rm DATA}}(\alpha)$
is interpreted as the CL on $\alpha$.
Figure~\ref{fig:alpha} shows $1-{\rm CL}$ for
$\alpha$ obtained from this method.

It is possible to obtain a prediction of $\alpha$ from indirect constraints by combining measurements of the CKM
matrix elements $|\vus|$, $|\vud|$, $|\vub|$, and $|\vcb|$, \CP\ violation in mixing from neutral kaons,
$\B$-$\Bbar$ mixing in $\B_d$ and $\B_s$ mesons and the measurement of $\sin 2\beta$ from $b \to
c\overline{c}s$ decays.
The indirect constraint on $\alpha$ from the UTfit~\cite{utfitter} and CKMfitter~\cite{ckmfitter} groups are
\utfitalpha\ and \ckmfitalpha\ respectively.  
Recent calculations using QCD Factorization predict $\alpha = (85.6^{+7.4}_{-7.3})^\circ$~\cite{ref:benekeqcdfact}.
Selecting the solution closest to the CKM
combined fit average we find $\alpha =
\measuredalpha$, where the error is
dominated by $|\delta\alpha|$ which is
$\measureddeltaalpha$ at $68\%$ CL.
The constraint obtained on $\alpha$ is worse than the previous
result in Ref.~\cite{babarrhoprhomr14} for the following reasons: (i) the
central value of the branching fraction for $\Bp \to \rho^+\rho^0$~\cite{babarrhoprhoz} is
smaller than the value previously used~\cite{babarvvpaper,bellerhorho0}, 
and (ii) the results of the latest
search for \Bztorhozrhoz\ show evidence for a signal~\cite{babarrhozrhoz}, 
while the previous constraint on $\alpha$ used the central value corresponding to the
upper limit available at that time~\cite{babarrhozrhozr14}.
Both of these factors lead to an increase in the penguin contribution
to the total uncertainty on $\alpha$ when using an isospin analysis.

A future extension to the \su{2} isospin method to measure $\alpha$ described here will be possible when
there are sufficient statistics to perform a time-dependent analysis
of \Bztorhozrhoz.  The parameters \slong\ and \clong\ measured in \Bztorhozrhoz\
can be incorporated into the isospin analysis, allowing one to 
over-constrain the isospin triangles of Eq. (\ref{eq:iaone}) and Eq. (\ref{eq:iatwo}).

\subsection{\boldmath Flavor \su{3} analysis}

There has been progress on constraining the penguin contribution in \Bztorhoprhom\ using
a flavor \su{3} based approach and experimental constraints from $B^+\to K^{*0}\rho^{+}$~\cite{ref:benekesuthree}.
The amplitude for \Bztorhoprhom\ can be written as
\begin{eqnarray}
{\cal A}(\Bztorhoprhom) = Te^{i\gamma} + P e^{i\deltatp},
\end{eqnarray}
where $T$ and $P$ are the magnitudes of tree and penguin amplitudes, $\gamma$ is 
the phase of \vub, and $\deltatp$ is the strong phase difference
between $T$ and $P$. The approach of Ref.~\cite{ref:benekesuthree} relates the penguin amplitude in
\Bztorhoprhom\ to the amplitude of the penguin decay $B^+\to K^{*0}\rho^{+}$ giving
three relations:
\begin{widetext}
\begin{eqnarray}
  \clong &=&\frac{2 r \sin\deltatp \sin(\beta + \alpha)}{ 1 - 2r \cos\deltatp \cos(\beta + \alpha) + r^2}, \label{eq:constraint1}\\
  \slong &=&\frac{\sin2\alpha + 2r \cos\deltatp\sin(\beta-\alpha) - r^2 \sin2\beta}{1 - 2r \cos\deltatp\cos(\beta+\alpha) + r^2},\nonumber \\
\label{eq:constraint2} \\
  \left( \frac {|\vcd|f_\rho} {|\vcs| f_{K^*}} \right)^2 \frac{ \Gamma_{\mathrm L} (\Bch \to K^{*0}\rho^+) } {\Gamma_{\mathrm L} (\Bztorhoprhom)}
    &=& \frac {Fr^2}{1 - 2r \cos\deltatp \cos(\beta+\alpha) + r^2},\label{eq:constraint3}
\end{eqnarray}
\end{widetext}
with three unknowns, $\alpha$, $r$, and $\deltatp$. The parameter $r=|P/T|$, and 
$\beta$ is the phase of \vtd.  We use the value of $\beta$ obtained
from $b\to c\overline{c}s$ decays~\cite{babar_sin2beta_2002,belle_sin2beta_2002}. The 
\CP\ averaged decay rates of the longitudinal components of $\Bch \to K^{*0}\rho^+$ 
and $\Bztorhoprhom$ ($\Gamma_{\mathrm L}$) are related 
by the squared ratio of CKM matrix elements and decay constants ($f_{i}$, where $\rho$ or $K^*$) as shown in Eq. (\ref{eq:constraint3}).
The factor $F$ accounts for additional sources of \su{3}\ breaking not described by the decay constants.

\begin{figure}[!ht]
 \begin{center}
 \resizebox{8cm}{!}{
 \includegraphics{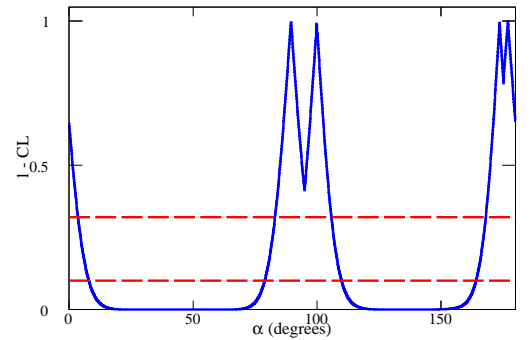}
}
\caption{Confidence level on $\alpha$ obtained from the flavor \su{3} approach.
The horizontal dashed lines correspond to the 68\% (top) and 90\% (bottom) 
CL intervals. The constraint shown in this figure uses some of the same 
inputs as the \su{2} isospin constraint of Fig.~\ref{fig:alpha}, and the two 
can not be averaged.} \label{fig:alphasu3}
 \end{center}
\end{figure}

The assumptions used in this approach are:
\begin{itemize}
\item The amplitude in the penguin dominated decay $B^+ \to K^{*0}\rho^+$ is 
      related to the penguin amplitude in \Bztorhoprhom,
\item The dominant \su{3} breaking correction accounted for by $F$ is the
      neglect of annihilation diagrams in the $B^+ \to K^{*0}\rho^+$ decay.  We use the same
      value of $F$ as Ref.~\cite{ref:benekesuthree}; $F=0.9 \pm 0.6$.
\end{itemize}
The result is a constraint on $\alpha$ with theoretical uncertainty from penguin 
contributions of \publishedmodelerroronalpha\ which represents
a stronger constraint on the unitarity triangle than provided by 
the isospin analysis constraint.  Figure~\ref{fig:alphasu3}
shows $1-{\rm CL}$ for $\alpha$ obtained with this method.  Selecting the solution closest to the CKM
combined fit average~\cite{ckmfitter,utfitter} we find $\alpha =
\measuredalphamodelnodeltaconstraint$, where the error is dominated by the experimental uncertainty on \slong\ and \clong.
The strong phase difference $\deltatp$ is only weakly constrained, 
and $r=\modelboundonr$.
The QCD factorization approach in Refs.~\cite{Beneke:1999br,Beneke:2000ry,Beneke:2001ev} predicts that \deltatp\ is small, and so we require 
$|\deltatp|<90^\circ$ as in Ref.~\cite{ref:benekesuthree}. We find $\alpha = \measuredalphamodel$.
These constraints are in agreement with the prediction for $\alpha$ from  
UTfit~\cite{utfitter} and CKMfitter~\cite{ckmfitter}, these predictions are \utfitalpha\ and \ckmfitalpha, respectively.

%\input{conclusions}
%-----------------------
% Conclusions
%-----------------------
\section{\boldmath Conclusions\label{sec:conclusions}}
\label{sec:conclusion}
We report the measurement of the branching fraction, \ptrue, and CP violation parameters, \slong\ and \clong,
for the decay \Bztorhoprhom\ using a data sample of \nbb.  We 
obtain the following results:
\begin{eqnarray}
\br(\Bztorhoprhom)&=& \measuredbf, \nonumber\\
\ptrue            &=& \measuredfl, \nonumber\\
{\slong}          &=& \measuredslong, \nonumber \\
\clong            &=& \measuredclong.\nonumber
\end{eqnarray}
Using these results, and experimental knowledge of the other $\B\to\rho\rho$
final states, we perform an isospin analysis to obtain a measurement of the CKM
angle $\alpha$.  The confidence level distribution for $\alpha$ is 
shown in Fig.~\ref{fig:alpha}.  The solution obtained that is compatible 
with the results of SM-based fits of existing data is $\alpha=\measuredalpha$.  The uncertainty on the measurement of 
$\alpha$ from the isospin analysis is dominated by penguin pollution.  If one uses 
the flavor \su{3} approach described in the text to constrain $\alpha$, one obtains the 
constraint $\alpha = \measuredalphamodelnodeltaconstraint$. 

%-----------------------
% Acknowledgements
%-----------------------
% Input the pubboard acknowledgements file
\section{Acknowledgments}
We are grateful for the 
extraordinary contributions of our \pep2\ colleagues in
achieving the excellent luminosity and machine conditions
that have made this work possible.
The success of this project also relies critically on the 
expertise and dedication of the computing organizations that 
support \babar.
The collaborating institutions wish to thank 
SLAC for its support and the kind hospitality extended to them. 
This work is supported by the
US Department of Energy
and National Science Foundation, the
Natural Sciences and Engineering Research Council (Canada),
the Commissariat \`a l'Energie Atomique and
Institut National de Physique Nucl\'eaire et de Physique des Particules
(France), the
Bundesministerium f\"ur Bildung und Forschung and
Deutsche Forschungsgemeinschaft
(Germany), the
Istituto Nazionale di Fisica Nucleare (Italy),
the Foundation for Fundamental Research on Matter (The Netherlands),
the Research Council of Norway, the
Ministry of Science and Technology of the Russian Federation, 
Ministerio de Educaci\'on y Ciencia (Spain), and the
Science and Technology Facilities Council (United Kingdom).
Individuals have received support from 
the Marie-Curie IEF program (European Union) and
the A. P. Sloan Foundation.

\appendix
%\input{interference}
%------------------------------------
% Summary of the toy MC technique used to estimate the interference 
% systematic uncertainty on the signal observables
%------------------------------------
\section*{\boldmath APPENDIX: Interference of \Bztorhoprhom\ with other $\Bz\to 4\pi$ modes}
\label{sec:interference}

An extensive study of the \B\ backgrounds, and the associated
systematic error, has been presented earlier in this paper. 
A number of \B backgrounds decay into the same final state as the signal. 
These are:
\begin{itemize}
\item $\Bz\to a_1^\pm\pi^\mp, a_1^\pm\to\rho^\pm\piz$,
\item $\Bz\to\rho^\mp\pi^\pm\piz$, non resonant,
\item $\Bz\to\pip\pim\piz\piz$, non resonant.
\end{itemize}
The systematic error associated with these modes was evaluated by
propagating the uncertainty on the branching ratio and \CP
asymmetries to the final measurements, with the appropriate
description of acceptance, SCF, resolutions and other
reconstruction biases (see Section~\ref{sec:Systematics}). 
The likelihood described in Section~\ref{sec:maximum}
does not account for possible interference between
$\Bz\to\rhop\rhom$ and other $\Bz\to 4\pi$ final states.

Acceptance, SCF, and other reconstruction biases are not
taken into account in this study, and instead, the $\rhop\rhom$
measurements are averaged in a perfect region of the phase space
referred to as the ``$\rhop\rhom$ band'', on which the analysis
technique is assumed to have a uniform sensitivity. This ``$\rhop\rhom$ band''
 corresponds to the kinematic selection of $\rho^\pm$  
(see Section~\ref{sec:rhomasshelicity}): $0.5 < m_{\pi^\pm\piz} < 1.0\gevcc$ and 
 $-0.9 < \cos\theta_i < 0.98$.

The remainder of this Appendix describes the final state wave function,
decay amplitudes, effective \CP\ asymmetries and the 
estimate of the systematic uncertainty from neglecting interference between
the signal and other $\Bz\to 4\pi$ final states.

\subsection{The final state wave functions}
\label{sec:final-state-wave}

This section summarizes the kinematic dependence of the decay
amplitudes.

\subsubsection{Lorentz invariant phase space}
\label{sec:lorentz-invar-phase}

The four-particle final state can be described completely in terms of
the following five variables:
\begin{itemize}
\item $\mvone$ and $\mvtwo$: the masses of the $\pip\piz$ and
  $\pim\piz$ pairs
\item $\coshelone$ and $\cosheltwo$: the angle between the
  $\pi^\pm\piz$ pair line of flight (as seen in the \B frame) and the
  $\piz$ in the pion pair frame.
\item $\phi$: the angle between the two planes defined by the
  $\pip\piz$ and $\pim\piz$ pairs in the \B frame.
\end{itemize}
We write each phase space dependent quantity $d{\Phi_4}$ as a function
of these five variables. These are:
\begin{equation}
  \label{eq:1}
  d{\Phi_4} \propto \frac{P^{\pip\piz}_{\B}}{M_{\B}}
  \frac{P^{\piz}_{\pip\piz}}{\mvone}
  \frac{P^{\piz}_{\pim\piz}}{\mvtwo}
  d\mvone^2 d\mvtwo^2 d\coshelone d\cosheltwo d\phi,
\end{equation}
where $P^{\pip\piz}_{\B}$ denotes the momentum of the pion pair in the
$\B$ frame and $M_{\B}$ is the mass of the $\B$ (and similarly for the
other quantities in this expression).

\subsubsection{Kinematic dependence of $\rhop\rhom$}
\label{sec:kinem-depend-rhoprh}

We assume that the longitudinal polarization of $\rhop\rhom$ is dominant,
so the kinematic dependence of the $\rhop\rhom$ wave function is:
\begin{equation}
  \label{eq:2}
  \begin{split}
    f(\rhop\rhom) &\propto BW(\mvone) BW(\mvtwo) Y_1^0(\theta_1)
    Y_1^0(\theta_2)\\
    &\propto BW(\mvone) BW(\mvtwo) \cos(\theta_1) \cos(\theta_2),
  \end{split}
\end{equation}
where $BW$ denotes a Breit-Wigner function (the simple
non-relativistic form is assumed), and $Y_l^m$ are the spherical
harmonics.

\subsubsection{Kinematic dependence of $a_1\pi$}
\label{sec:kinem-depend-a_1pi}

\begin{figure}[htbp]
  \centering
  \includegraphics[width=0.4\textwidth]{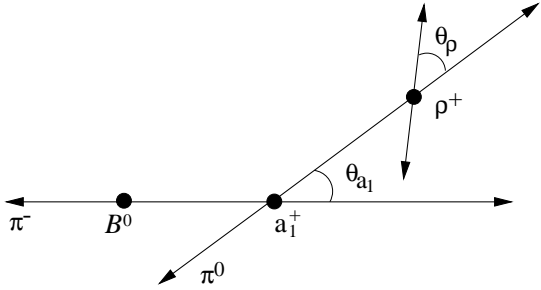}
  \caption{Definition of the variables $\theta_\rho$ and $\theta_{a_1}$ entering the $a_1\pi$ kinematic
  dependence.}
  \label{fig:a1pidef}
\end{figure}

In order to write the wave function of $a_1^+\pim$, we need to define
the following three variables, which are functions of the variables
found in part 1a of this Appendix:
\begin{itemize}
\item $\theta_\rho$: the angle between the \piz and the $\rho$ line of
  flight (as seen from the $a_1$) in the $\rho$ frame,
\item $\theta_{a_1}$: the angle between the $\rho$ and $a_1$ line of
  flight (as seen from the $\B$) in the $a_1$ frame,
\item $\phi_{a_1}$: the angle between the two decay planes of the $a_1$
  and $\rho$ mesons, in the $a_1$ frame.
\end{itemize}
A schematic view of $\Bz\to a_1^\pm\pi^\mp$ showing $\theta_\rho$ and $\theta_{a_1}$ is given in
Fig.~\ref{fig:a1pidef}.

The $a_1$ has the possibility to decay both into an $S$ wave and a $D$
wave. We assume that the $S$ wave contribution is  dominant, so the 
kinematic dependence of $a_1\pi$ is:
\begin{equation}
  \label{eq:3}
  \begin{split}
    f(a_1\pi) \propto & BW(a_1) BW(\rho) \\
    &\bigg{\{} d^1_{0,1}(\theta_{a_1}) Y^1_1(\theta_\rho, \phi_{a_1})\\
    & + d^1_{0,0}(\theta_{a_1}) Y^0_1(\theta_\rho)\\
    & + d^1_{0,-1}(\theta_{a_1}) Y^{-1}_1(\theta_\rho, \phi_{a_1})\bigg{\}},
  \end{split}
\end{equation}
where the $d^{j}_{m, m^\prime}$ are rotation matrices.
A non-relativistic Breit-Wigner is assumed for $BW(\rho)$, and a
relativistic form is used for $BW(a_1)$.

\subsubsection{Kinematic dependence of the non-resonant
  contributions}
\label{sec:kinem-depend-non}
The kinematic dependence of the non-resonant contributions can be
written as:
\begin{equation}
  \label{eq:4}
  f(\rho^\mp\pi^\pm\piz) \propto BW(\rho_1) Y_1^0(\theta_1),
\end{equation}
for $\rho^\mp\pi^\pm\piz$, and:
\begin{equation}
  \label{eq:5}
  f(\pip\pim\piz\piz) = \text{constant},
\end{equation}
for the non-resonant 4-pion final state.

We note that the spherical harmonics entering the $\rhop\rhom$
amplitudes found in Eq. (\ref{eq:2}) differ from those found in these
two amplitudes: this means that these wave functions are orthogonal,
and after the integration over the helicity angles, the interfering
terms $f(\rhop\rhom) f^\star(\rho^\mp\pi^\pm\piz)$ and $f(\rhop\rhom)
f^\star(\pip\pim\piz\piz)$ will cancel completely if the integration
is done over the whole phase space.
The ``\rhop\rhom band'' used in this analysis is slightly asymmetric in
helicity angle, and as a result we can expect a small deviation from zero for these
integrals.

\subsubsection{Symmetrization of the wave functions}
\label{sec:symm-wave-funct}

The final state is made of four bosons, two of which are identical (the two
\piz). Thus each of the above wave functions needs to be symmetrized,
that is to say that the amplitude:
\begin{equation}
  \label{eq:6}
  f(p_{\pip}, p_{\piz}^1, p_{\pim}, p_{\piz}^2),
\end{equation}
where $p$ denotes the four-momentum of one of the pions, needs to be
replaced by:
\begin{eqnarray}
  \label{eq:7}
  \tilde{f}(p_{\pip}, p_{\piz}^1, p_{\pim}, p_{\piz}^2)
 & =& \frac{1}{\sqrt{2}} \bigg{\{} f(p_{\pip}, p_{\piz}^1, p_{\pim}, p_{\piz}^2)
\nonumber
\\
 && +   f(p_{\pip}, p_{\piz}^2, p_{\pim}, p_{\piz}^1)\bigg{\}}.
\end{eqnarray}
In the rest of this study, each of the wave functions are symmetrized by
$\piz$ meson exchange. 

\subsection{Decay amplitudes}
\label{sec:compl-decay-ampl}

For a particular contribution $\Bz\to x$, the total decay
amplitude $\mathcal{A}(\Bz\to x)$ is the product of a phase space
independent amplitude $A(\Bz\to x)$ and one of the above kinematic
dependences $\tilde{f}(x)$:
\begin{equation}
  \label{eq:8}
  \mathcal{A}(\Bz\to x) = A(\Bz\to x) \tilde{f}(x).
\end{equation}

This section describes how we calculate $A(\Bz\to x)$.

\subsubsection{Amplitude of $\B\to\rhop\rhom$}
\label{sec:compl-ampl-btorh}

For $\B\to\rhop\rhom$, two amplitudes need to be known:
$A(\Bz\to\rhop\rhom)$ and $A(\Bzb\to\rhop\rhom)$, corresponding to 4
degrees of freedom. We choose the following parameterization:
\begin{itemize}
\item The branching ratio ${\cal B}(\B\to\rhop\rhom)$,
\item The \CP-violating asymmetries $\clong(\rhop\rhom)$ and $\slong(\rhop\rhom)$,
\item An overall phase $\delta(\rhop\rhom)$ which may be set to zero
  in the appropriate phase convention.
\end{itemize}
The parameters ${\cal B}(\B\to\rhop\rhom)$, $\clong(\rhop\rhom)$ and
$\slong(\rhop\rhom)$ are set to the values reported here
(see Section~\ref{sec:conclusion}).

\subsubsection{Amplitude of $\B\to a_1\pi$}
\label{sec:compl-ampl-bto}

Four amplitudes describe the $\B\to a_1\pi$ contributions,
corresponding to a total of 8 degrees of freedom: $A(\Bz\to
a_1^+\pim)$, $A(\Bz\to a_1^-\pip)$, $A(\Bzb\to a_1^+\pim)$ and
$A(\Bzb\to a_1^-\pip)$. We choose a parameterization similar to that
of the time dependent $a_1\pi$ analysis \cite{a1piTD}, for which parameters have
direct physical meaning:
\begin{itemize}
\item $\BR(\Bz \to a_1\pi)$ is the total branching ratio,

\item $C(a_1\pi)+\delC(a_1\pi) = C^+(a_1\pi) =
  \frac{1-|\lambda_{\CP} (a_1^+\pim)|^2}{1+|\lambda_{\CP} (a_1^+\pim)|^2}$ where
  $\lambda_{\CP} (a_1^+\pim) = \frac{q}{p}\frac{A(\Bzb\to
    a_1^+\pim)}{A(\Bz\to a_1^+\pim)}$,

\item $C(a_1\pi)-\delC(a_1\pi) = C^-(a_1\pi) =
  \frac{1-|\lambda_{\CP} (a_1^-\pip)|^2}{1+|\lambda_{\CP} (a_1^-\pip)|^2}$ where
  $\lambda_{\CP} (a_1^-\pip) = \frac{q}{p}\frac{A(\Bzb\to
    a_1^-\pip)}{A(\Bz\to a_1^-\pip)}$, $C(a_1\pi)$ is a direct
  \CP-violating asymmetry and $\delC(a_1\pi)$ is \CP-conserving,

\item $S(a_1\pi)+\delS(a_1\pi) = S^+(a_1\pi) = \frac{2{\cal I}m
    \{\lambda_{\CP} (a_1^+\pim)\}}{1+|\lambda_{\CP} (a_1^+\pim)|^2}$,

\item $S(a_1\pi)-\delS(a_1\pi) = S^-(a_1\pi) = \frac{2{\cal I}m
    \{\lambda_{\CP} (a_1^-\pip)\}}{1+|\lambda_{\CP} (a_1^-\pip)|^2}$, $S(a_1\pi)$ is a
    \CP-violating asymmetry in the interference between decay and
    mixing, and $\delS(a_1\pi)$ is \CP conserving,

\item $\Acp(a_1\pi) = \frac{|A^{+-}|^2 + |\overline{A}^{+-}|^2 
  - |A^{-+}|^2 - |\overline{A}^{-+}|^2}
   {|A^{+-}|^2 + |\overline{A}^{+-}|^2 
  - |A^{-+}|^2 - |\overline{A}^{-+}|^2}$ 
  where $A^{\pm\mp}=A(\Bz\to a_1^\pm\pimp)$, and $\overline{A}^{\pm\mp}$ 
  is its complex conjugate.  These are also direct \CP violating asymmetries,

\item Two overall phases of the final states $\delta(a_1^+\pim)$ and
  $\delta(a_1^-\pip)$.
\end{itemize}

For $a_1\pi$ we assume that all of these parameters are constant over
the whole of phase space.

\smallskip

\subsection{Effective \CP asymmetries}
\label{sec:effect-cp-asymm}

Considering the two final states $\rhop\rhom$ and $a_1\pi$, the 
phase-space dependent four-pion decay amplitude is:
\begin{eqnarray}
\begin{split}
    \mathcal{A}(\Bz\to 4\pi) 
    &= \mathcal{A}(\Bz\to\rhop\rhom) +  \mathcal{A}(\Bz\to a_1^+\pim)  \nonumber \\
    &\quad + \mathcal{A}(\Bz\to a_1^-\pip),\nonumber \\
    &= A(\Bz\to\rhop\rhom)f(\rhop\rhom) \\
    &\quad + A(\Bz\to a_1^+\pim)f(a_1^+\pim) \nonumber \\
    &\quad + A(\Bz\to a_1^-\pip)f(a_1^-\pip),
 \label{eq:9}
\end{split}
\end{eqnarray}
with a similar expression for $\mathcal{A}(\Bzb\to 4\pi)$. Similar equations 
exist for $\rho\pi\pi$ and $4\pi$ final states.

We can define the phase-space dependent quantity:
\begin{equation}
  \label{eq:10}
  \lambda_{4\pi}(\Phi_4) = \frac{q}{p} \frac{\mathcal{A}(\Bzb\to
  4\pi)}{\mathcal{A}(\Bz\to 4\pi)},
\end{equation}
where $\lambda_{4\pi}(\Phi_4)$ is a function of \mvone,
\mvtwo, \coshelone, \cosheltwo, and $\phi$. Thus
define the phase-space dependent \CP-violating asymmetries:
\begin{align}
  C_{4\pi}(\Phi_4) &=
  \frac{1-|\lambda_{4\pi}(\Phi_4)|^2}{1+|\lambda_{4\pi}(\Phi_4)|^2}, \\
  S_{4\pi}(\Phi_4) &=
  \frac{2{\cal I}m (\lambda_{4\pi}(\Phi_4))}{1+|\lambda_{4\pi}(\Phi_4)|^2},
\end{align}
that are analogous to Eq. (\ref{eq:slong}) and Eq. (\ref{eq:clong}).

In principle it is possible to measure the value of 
$C_{4\pi}(\Phi_4)$ and $S_{4\pi}(\Phi_4)$ at each point of
 phase space, however this would be experimentally challenging.
Instead, this measurement uses integrated values
$C_{\text{eff}}$ and $S_{\text{eff}}$, which can be seen as the
averages of $C$ and $S$ across the previously defined ``$\rhop\rhom$ band'', 
weighted over the number of events found at each location
of the phase space which corresponds to:

\begin{widetext} 
\begin{eqnarray}
  C_{\text{eff}} &=& \frac{\int_{\text{$\rho\rho$ band}}
    C_{4\pi}(\Phi_4) (|\mathcal{A}(\Bz\to 4\pi)|^2 +
    |\mathcal{A}(\Bzb\to 4\pi)|^2) d\Phi_4}
  {\int_{\text{$\rho\rho$ band}} (|\mathcal{A}(\Bz\to 4\pi)|^2 +
    |\mathcal{A}(\Bzb\to 4\pi)|^2) d\Phi_4}, \nonumber \\
  S_{\text{eff}} &=& \frac{\int_{\text{$\rho\rho$ band}}
    S_{4\pi}(\Phi_4) (|\mathcal{A}(\Bz\to 4\pi)|^2 +
    |\mathcal{A}(\Bzb\to 4\pi)|^2) d\Phi_4}
  {\int_{\text{$\rho\rho$ band}} (|\mathcal{A}(\Bz\to 4\pi)|^2 +
    |\mathcal{A}(\Bzb\to 4\pi)|^2) d\Phi_4}.
\end{eqnarray}
\end{widetext} 

We can also define the values of $C$ and $S$ that would be observed in the
absence of interference effects, denoted by $C^{ \not 
 \mathcal{I}}_{4\pi}(\Phi_4)$ and $S^{\not
  \mathcal{I}}_{4\pi}(\Phi_4)$, and similarly the integrated
values across the ``$\rhop\rhom$ band'' $C^{\not
  \mathcal{I}}_{\text{eff}}$ and $S^{\not \mathcal{I}}_{\text{eff}}$.
Finally, for the evaluation of the systematic error, we are really
interested in the biases on $C$ and $S$ purely due to interference 
which are the two differences 
$ \delta C = C_{\text{eff}} - C^{\not \mathcal{I}}_{\text{eff}}$
and $\delta S= S_{\text{eff}} - S^{\not \mathcal{I}}_{\text{eff}}$.
We adopt the same strategy for the branching fraction $\BR(\rhop\rhom)$
by defining $ \delta \BR =  \BR_{\text{eff}} - \BR^{\not \mathcal{I}}_{\text{eff}}$
in a similar way.

\subsection{The systematic error}
\label{sec:systematic-error}

As the 8 parameters describing the amplitude $A(\Bz(\Bzb) \to a_1\pi)$ are
not yet measured, to estimate the systematic error on $\rhop\rhom$ parameters, 
we need to propagate the uncertainty on these 8 parameters. For this,
we use a MC technique. For each experiment, we generate  
the \CP\ parameters $C(a_1\pi)$,
$\delC(a_1\pi)$, $S(a_1\pi)$, $\delS(a_1\pi)$, $\Acp(a_1\pi)$, the
phases $\delta(a_1^+\pim)$ and $\delta(a_1^-\pip)$,
and the branching fraction $\BR(a_1\pi)$. 
The values used to generate the $a_1\pi$ contribution have Gaussian PDFs 
for measured quantities (with mean and width
corresponding to the measurement), and uniform PDFs for unmeasured 
quantities (sampled over the allowed values).
Then, we estimate the value of $\delta C$ and $\delta S$ 
 for each generated value of the $a_1\pi$ parameters. The systematic 
error are taken as the RMS of the $\delta C$, $\delta S$ and  $\delta \BR$
distributions.

For this study, the quantity $\BR(a_1\pi)$ is fixed to the value measured 
in~\cite{ref:bchn}, $\BR(\Bz\to a_1^+\pim, a_1^+\to\rhop\piz) = (16.6\pm 2.4) \times 10^{-6}$ 
and the \CP\ parameters are obtained from~\cite{a1piTD}: 
$C(a_1\pi) = -0.10 \pm 0.17 $,
$\delC(a_1\pi)= 0.26 \pm 0.17$, $S(a_1\pi) = 0.37 \pm 22$, 
$\delS(a_1\pi) = -0.14 \pm 0.22$ and $\Acp(a_1\pi)= -0.07 \pm 0.07$.   

We estimate the effect of interference between signal and either $\rho^\mp\pi^\pm\piz$ or $\pip\pim\piz\piz$
in a similar way, where we assume the branching fraction for these two non-resonant backgrounds
is $(5.7\pm 13.0) \times 10^{-6}$. This value is obtained by modeling the
$\rho^\mp\pi^\pm\piz$ component in the likelihood fit and by fitting  a free yield
to the data, and we assume that $\rho^\mp\pi^\pm\piz$ and $\pip\pim\piz\piz$ have
the same branching fraction.

We summarize the systematic uncertainty obtained for the different $4\pi$ final 
states considered here in Table~\ref{tab:interfsysttotal}. The total uncertainty on
\BR\ is used to calculate the quoted systematic uncertainty on \ptrue\ from this source. We conclude that
interference between signal and other $4\pi$ final
states has a negligible impact on the measurement.
\begin{table}[htbp]
  \centering
  \caption{Systematic error coming from interference effects due to
    three contaminating amplitudes.}
  \label{tab:interfsysttotal}
  \begin{tabular}{l|ccc}
    \hline\hline\noalign{\vskip1pt}
    Contribution & $\sigma(C)$ & $\sigma(S)$ & $\frac{\sigma(\BR)}{\BR}$ \\
    \noalign{\vskip1pt} \hline
    $a_1\pi$ & 0.009 & 0.009 & 2.5\% \\
    $\rho\pi\pi$ non resonant & 0.010 & 0.010 & 1.1\% \\
    $4\pi$ non resonant & 0.002 & 0.002 & 0.3\% \\
    \hline
    Total & 0.014 & 0.014 & 2.7\% \\
    \hline\hline
  \end{tabular}
\end{table}

\newpage   

\bibliographystyle{unsrt}
\bibliography{biblio}

\begin{thebibliography}{10}

\bibitem{christenson}
J.~H. Christenson, J.~W. Cronin, V.~L. Fitch, and R.~Turlay.
\newblock {\em \jprl{13}}, 138 (1964).

\bibitem{asakharov}
A.~D. Sakharov.
\newblock {\em Pisma Zh. Eksp. Teor. Fiz.}, {\bf{5}}, 32 (1967); Traduction
  JETP Lett {\bf{5}}, 24 (1967).

\bibitem{CKM1}
N.~Cabibbo.
\newblock {\em \jprl{10}}, 531 (1963).

\bibitem{CKM2}
M.~Kobayashi and T.~Maskawa.
\newblock {\em \progtp{49}}, 652 (1973).

\bibitem{wolfenstein}
L.~Wolfenstein.
\newblock {\em \jprl{51}}, 1945 (1983).

\bibitem{buras}
A.~J. Buras, M.~E. Lautenbacher, and G.~Ostermaier.
\newblock {\em \jprd{50}}, 3433 (1994).

\bibitem{babar_sin2beta_2002}
B.~Aubert \etal.~(\babar\ Collaboration).
\newblock {\em \jprl{89}}, 201802 (2002).

\bibitem{belle_sin2beta_2002}
K.~Abe \etal. (Belle~Collaboration).
\newblock {\em \jprd{66}}, 071102 (2002).

\bibitem{gavela}
M.~B. Gavela.
\newblock {\em Mod. Phys. Lett. A {\bf{9}}}, 795 (2002).

\bibitem{huet}
P.~Huet and E.~Sather.
\newblock {\em \jprd{51}}, 379 (1995).

\bibitem{babar_sin2beta}
B.~Aubert \etal.~(\babar\ Collaboration).
\newblock hep-ex/0703021.

\bibitem{belle_sin2beta}
K.-F.~Chen \etal. (Belle~Collaboration).
\newblock {\em \jprl{98}}, 031802 (2007).

\bibitem{ref:ciuchini1995}
M.~Ciuchini, E.~Franco, G.~Martinelli, L.~Reina, and L.~Silvestrini.
\newblock {\em \zpc{68}}, 239 (1995).

\bibitem{ckmfitter}
CKMfitter Group.
\newblock {\em Eur. Phys. J. C {\bf 41}}, 1-131 (2005).
\newblock Updated results and plots available at: http://ckmfitter.in2p3.fr.

\bibitem{utfitter}
UTfit Collaboration.
\newblock {\em JHEP {\bf 0507}}, 028 (2005).
\newblock Updated results and plots available at: http://www.utfit.org.

\bibitem{bigiandsanda}
I.~Bigi and A.~Sanda.
\newblock {\em \rm{\underline{CP Violation}}}.
\newblock Cambridge University Press, 1999.

\bibitem{brancolavouraandsilva}
G.~Branco, L.~Lavoura, and J.~Silva.
\newblock {\em \rm{\underline{CP Violation}}}.
\newblock Oxford University Press, 1999.

\bibitem{conj}
Unless explicitly stated, charge conjugation is implied throughout.

\bibitem{bevan2006}
A.~Bevan.
\newblock {\em Mod. Phys. Lett. A {\bf 21}}, 305 (2006).

\bibitem{gronaulondon}
M.~Gronau and D.~London.
\newblock {\em \jprl{65}}, 3381 (1990).

\bibitem{babarrhoprhoz}
B.~Aubert \etal.~(\babar\ Collaboration).
\newblock {\em \jprl{97}}, 261801 (2006).

\bibitem{bellerhorho0}
J.~Zhang \etal. (Belle~Collaboration).
\newblock {\em \jprl{91}}, 221801 (2003).

\bibitem{babarrhozrhoz}
B.~Aubert \etal.~(\babar\ Collaboration).
\newblock {\em \jprl{98}}, 111801 (2007).

\bibitem{ref:benekesuthree}
M.~Beneke, M.~Gronau, J.~Rohrer, and M.~Spranger.
\newblock {\em \plb{638}}, 68 (2006).

\bibitem{ref:zupanckm}
See, for example, J.~Zupan hep-ph/0701004, and references therein.

\bibitem{babarrhoprhomr14}
B.~Aubert \etal.~(\babar\ Collaboration).
\newblock {\em \jprl{95}}, 041805 (2005).

\bibitem{bellerhoprhom}
A.~Somov \etal. (Belle~Collaboration).
\newblock {\em \jprl{96}}, 171801 (2006).

\bibitem{pdg2006}
W.-M.~Yao \etal. (Particle Data~Group).
\newblock {\em \jpg{33}}, 1 (2006).

\bibitem{ref:lambda}
See, for example, D.~Kirkby and Y.~Nir, Section 12, in Ref.~\cite{pdg2006}.

\bibitem{babarrhoprhomprlr12}
B.~Aubert \etal.~(\babar\ Collaboration).
\newblock {\em \jprl{93}}, 231801 (2004).

\bibitem{bellerhoprhomupdate}
A.~Somov \etal. (Belle~Collaboration).
\newblock hep-ex/0702009.

\bibitem{ref:pepcdr}
{P}{E}{P}-{I}{I} {C}onceptual {D}esign {R}eport.
\newblock {\em SLAC-R-418}, (1993).

\bibitem{babar_nim}
B.~Aubert \etal.~(\babar\ Collaboration).
\newblock {\em \nima{479}}, 1 (2002).

\bibitem{ref:lsta}
G.~Benelli, K.~Honscheid, E.~A. Lewis, J.~J. Regensburger, and D.~S. Smith.
\newblock {\em Nuclear Science Symposium Conference Record, 2005 IEEE,
  {\bf{2}}}, 1145 (2005).

\bibitem{ref:lstb}
W.~Menges.
\newblock {\em Nuclear Science Symposium Conference Record, 2005 IEEE,
  {\bf{3}}}, 1470 (2005).

\bibitem{ref:lstc}
M.~R. Convery, P.~C. Kim, H.~P. Paar, C.~H. Rogers, R.~H. Schindler, S.~K.
  Swain, and C.~C. Young.
\newblock {\em \nima{556}}, 134 (2006).

\bibitem{ref:geant}
S.~Agostinelli \etal. (GEANT4~Collaboration).
\newblock {\em \nima{506}}, 250 (2003).

\bibitem{ref:lat}
A.~Drescher, B.~Grawe, B.~Hahn, B.~Ingelbach, U.~Matthiesen, H.~Scheck,
  J.~Spengler, and D.~Wegener.
\newblock {\em \nima{237}}, 464 (1985).

\bibitem{foxwolfram}
G.~C. Fox and S.~Wolfram.
\newblock {\em \jpl{41}}, 1581 (1978).

\bibitem{babarsin2betaprd}
B.~Aubert \etal.~(\babar\ Collaboration).
\newblock {\em \jprd{66}}, 032003 (2002).

\bibitem{cba}
M.~J. Oreglia.
\newblock Ph.D Thesis, SLAC-236, Appendix D, (1980).

\bibitem{cbb}
J.~E. Gaiser.
\newblock Ph.D Thesis, SLAC-255, Appendix F, (1982).

\bibitem{cbc}
T.~Skwarnicki.
\newblock Ph.D Thesis, DESY F31-86-02, Appendix E, (1986).

\bibitem{argus2}
H.~Albrecht \etal. (ARGUS~Collaboration).
\newblock {\em Phys Lett}, B241:278, (1990).

\bibitem{ref:bchh}
B.~Aubert \etal.~(\babar\ Collaboration).
\newblock {\em \jprd{72}}, 072003 (2005).

\bibitem{ref:bchi}
A.~Garmash \etal. (Belle~Collaboration).
\newblock {\em \jprl{96}}, 251803 (2006).

\bibitem{ref:bchc}
B.~Aubert \etal.~(\babar\ Collaboration).
\newblock hep-ex/0701035.

\bibitem{ref:bchd}
J.~Zhang \etal. (Belle~Collaboration).
\newblock {\em \jprl{94}}, 031801 (2005).

\bibitem{ref:bche}
CLEO Collaboration.
\newblock {\em \jprl{85}}, 2881 (2000).

\bibitem{ref:bchk}
B.~Aubert \etal.~(\babar\ Collaboration).
\newblock {\em \jprl{91}}, 201802 (2003).

\bibitem{ref:bchn}
B.~Aubert \etal.~(\babar\ Collaboration).
\newblock {\em \jprl{97}}, 051802 (2006).

\bibitem{ref:bchp}
B.~Aubert \etal.~(\babar\ Collaboration).
\newblock {\em \jprd{74}}, 031104 (2006).

\bibitem{ref:dcsd}
O.~Long, M.~Baak, R.~N. Cahn, and D.~Kirkby.
\newblock {\em \jprd{68}}, 034010 (2003).

\bibitem{electroweak_pengin_calculation}
See, for example, p73 of Ref.~\cite{ckmfitter}.

\bibitem{zupanewp}
M.~Gronau and J.~Zupan.
\newblock {\em \jprd{71}}, 074017 (2005).

\bibitem{gardner1999}
S.~Gardner.
\newblock hep-ph/9906269.

\bibitem{gardner2005}
S.~Gardner.
\newblock {\em \jprd{72}}, 034015 (2005).

\bibitem{ref:london2006}
F.~J. Botella, D.~London, and J.~P. Silva.
\newblock {\em \jprd{73}}, 071501 (2006).

\bibitem{falk}
A.~F. Falk, Z.~Ligeti, Y.~Nir, and H.~Quinn.
\newblock {\em \jprd{69}}, 011502 (2004).

\bibitem{FC98}
G.~Feldman and R.~Cousins.
\newblock {\em \jprd{57}}, 3873 (1998).

\bibitem{ref:benekeqcdfact}
M.~Beneke, J.~Rohrer, and D.~Yang.
\newblock hep-ph/0612290.

\bibitem{babarvvpaper}
B.~Aubert \etal.~(\babar\ Collaboration).
\newblock {\em \jprl{91}}, 171802 (2003).

\bibitem{babarrhozrhozr14}
B.~Aubert \etal.~(\babar\ Collaboration).
\newblock {\em \jprl{94}}, 131801 (2005).

\bibitem{Beneke:1999br}
M.~Beneke, G.~Buchalla, M.~Neubert, and C.~T. Sachrajda.
\newblock {\em \jprl{83}}, 1914 (1999).

\bibitem{Beneke:2000ry}
M.~Beneke, G.~Buchalla, M.~Neubert, and C.~T. Sachrajda.
\newblock {\em \npb{591}}, 313 (2000).

\bibitem{Beneke:2001ev}
M.~Beneke, G.~Buchalla, M.~Neubert, and C.~T. Sachrajda.
\newblock {\em \npb{606}}, 245 (2001).

\bibitem{a1piTD}
B.~Aubert \etal.~(\babar\ Collaboration).
\newblock {\em \jprl{98}}, 181803 (2007).

\end{thebibliography}

\end{document}